\documentclass[journal]{IEEEtran}
\usepackage{ bm,graphicx,amsmath,amssymb,multirow,graphicx}
\usepackage{algorithm,algpseudocode}

\usepackage{color, soul}

\begin{document}
\title{ Bayesian Hypothesis Test using Nonparametric Belief Propagation for Noisy Sparse Recovery }


\author{ Jaewook~Kang,~\IEEEmembership{Member,~IEEE,}
        Heung-No~Lee,~\IEEEmembership{Senior Member,~IEEE,}
        and~Kiseon~Kim,~\IEEEmembership{Senior Member,~IEEE,}
%


 \thanks{ The authors are with Department of Information and Communication, Gwangju Institute of Science and
 Technology, Gwangju, Republic of Korea
 (Email:\{jwkkang,heungno,kskim\}@gist.ac.kr)}

} \maketitle \setlength{\baselineskip}{1.0\baselineskip}
\begin{abstract}

This paper proposes a low-computational Bayesian algorithm for noisy
sparse recovery (NSR), called BHT-BP. In this framework, we consider
an LDPC-like measurement matrices which has a tree-structured
property, and additive white Gaussian noise. BHT-BP has a joint
detection-and-estimation structure consisting of a sparse support
detector and a nonzero estimator. The support detector is designed
under the criterion of the minimum detection error probability using
a nonparametric belief propagation (nBP) and composite binary
hypothesis tests. The nonzeros are estimated in the sense of linear
MMSE, where the support detection result is utilized. BHT-BP has its
strength in noise robust support detection, effectively removing
quantization errors caused by the uniform sampling-based nBP.
Therefore, in the NSR problems, BHT-BP has advantages over CS-BP
\cite{CS-BP2} which is an existing nBP algorithm, being comparable
to other recent CS solvers, in several aspects. In addition, we
examine impact of the minimum nonzero value of  sparse signals via
BHT-BP, on the basis of the results of
\cite{Wainwright1},\cite{Wainwright2},\cite{Fletcher}. Our empirical
result shows that  variation of $x_{min}$ is reflected to recovery
performance in the form of SNR shift.

\end{abstract}

\begin{keywords}
 Noisy sparse recovery, compressed sensing, nonparametric belief
 propagation,
 composite hypothesis testing,\\ joint detection-and-estimation
\end{keywords}

\section{Introduction}
\subsection{Background}
Robust reconstruction of sparse signals against measurement noise is
a key problem in real-world applications of compressed sensing (CS)
\cite{MRI}-\cite{ADC}. We refer to such signal recovery problems as
\emph{noisy sparse signal recovery} (NSR) problems.  The NSR
problems can be directly defined as an $l_0$-norm minimization
problem \cite{Donoho2},\cite{tropp}. Solving the $l_0$-norm task is
very limited in practice when the system size $(M,N)$ becomes large.
Therefore, several alternative solvers have been developed to relax
 computational cost of the $l_0$-norm task, such as $l_1$-norm
minimization solvers, \emph{e.g.}, Dantzig selector ($l_1$-DS)
\cite{candes2} and Lasso \cite{Lasso}, and greedy type algorithms,
\emph{e.g.}, OMP \cite{OMP} and COSAMP \cite{Cosamp}. Another
popular approach to the computational relaxation   is based on the
Bayesian philosophy \cite{SBL}-\cite{AMP1}. In the Bayesian
framework, the $l_0$-norm task is described as maximum a posteriori
(MAP) estimation problem, and sparse solution then is sought by
imposing a certain sparsifying prior probability density function
(PDF) with respect to the target signal \cite{Bruckstein}.

Recently,  Baysian solvers applying \emph{belief propagation} (BP)
have been introduced and caught attention as a low-computational
approach to handle the NSR problems in a large system setup
\cite{CS-BP2}-\cite{AMP1}. These BP-based solvers reduce
computational cost of the signal recovery by removing unnecessary
and duplicated computations using statistical dependency within the
linear system. Such BP solvers are also called message-passing
algorithms because their recovery behavior is well explained by
passing statistical messages over a tree-structured graph
representing the statistical dependency \cite{factor}.

For   implementation of BP, two approaches have been mainly
discussed according to  message representation methods: parametric
BP (pBP) \cite{BP-SBL}-\cite{AMP1},\cite{Freg},\cite{loopyBP} where
the BP-message is approximated to a Gaussian PDF; hence, only the
mean and variance are used for  message-passing, and  nonparametric
BP (nBP)
\cite{CS-BP2},\cite{SSP2012},\cite{non_para_BP}-\cite{nBP_sensor}
where the BP-message is represented by  samples of the corresponding
PDF. When the pBP approach is used, there are errors from the
Gaussian approximation; these errors decrease as problem size
$(N,M)$ increases. If the nBP approach is used, there is an
approximation error which generally depends upon the choice of
message sampling methods.

\subsection{Contribution}
In this paper, a low-computational Bayesian algorithm is developed
based on the nBP approach. We refer to the proposed algorithm as
\emph{Bayesian hypothesis test using nonparametric belief
propagation} (BHT-BP)\footnote{The MATLAB code of the proposed
algorithm is available at our webpage,
https://sites.google.com/site/jwkang10/}. Differently from the
pBP-based solvers, BHT-BP can precisely infer the multimodally
distributed BP-messages via an uniform sampling-based nBP.
Therefore, BHT-BP can be applied to any types of sparse signals in
the CS framework by adaptively choosing a signal prior PDF. In
addition, the proposed algorithm uses \emph{low-density parity-check
codes} \cite{Gallager} (LDPC)-like sparse measurement matrices as
works in \cite{CS-BP2},\cite{BP-SBL},\cite{SuPrEM}. Although such
sparse matrices perform worse than the dense matrices do in terms of
compressing capability in the CS framework, they can highly speed up
the generation of the CS measurements \cite{Wainwright1}.

Most CS algorithms to date for the NSR problems have been developed
under the auspices of signal estimation rather than support
detection.  However, recently studies have indicated that the
existing estimation-based algorithms, such as Lasso \cite{Lasso},
lead to a potentially large gap with respect to the theoretical
limit for the noisy support recovery
\cite{Wainwright2}-\cite{Fletcher}. Motivated by such theoretical
investigation, the proposed BHT-BP takes a joint
detection-and-estimation structure \cite{DE_info},\cite{Moustaki},
as shown in Fig.\ref{fig:Fig2-1}, which consists of a sparse support
detector and a nonzero estimator. The support detector uses uniform
sampling-based nBP and composite binary hypothesis tests to the CS
measurements $\mathbf{Z}$ at hand for the sparse support finding.
Given the detected support, the underdetermined CS problem is
reduced to an overdetermined problem. Then, the nonzero estimator is
applied under the criterion of \emph{linear minimum
mean-square-error} (LMMSE) \cite{KayII}. Then, let us state the
detailed novel points of the proposed algorithm. In the CS framework
considering reconstruction of a sparse signal
$\mathbf{X}\in\mathbb{R}^N$ from noisy measurements
$\mathbf{Z}\in\mathbb{R}^M$, BHT-BP is novel in terms of
\begin{enumerate}
\item Providing robust support detection against additive measurement noise based on \emph{the criterion of the minimum  detection error probability},
\item Removing MSE degradation caused by the message sampling  of the uniform sampling-based nBP using a joint detection-and-estimation
structure,
\item Handling sparse signals whose minimum nonzero value is regulated by a parameter $x_{\min}\geq 0$,   proposing a signal prior PDF for such
signals,
\item Providing fast sparse reconstruction with recovery complexity $\mathcal{O}(N\log N + KM)$ where $K$ is the signal sparsity.
\end{enumerate}

For the support detection of BHT-BP, we use a hypothesis-based
detector designed under the criterion of the minimum detection error
probability \cite{KayI}. BHT-BP represents the signal support using
a binary vector, scalarwisely applying the hypothesis testing  to
each binary element for the  support finding. This hypothesis test
is ``composite" because the likelihood for the test is associated
with the value of each scalar $X_i$. Therefore, we calculate the
likelihood under the Bayesian paradigm; then, the likelihood for the
test is a function of the signal prior and the marginal posterior of
$X_i$. This is the reason why we refer to our support detection as
\emph{Bayesian hypothesis test} (BHT) detection. BHT-BP has noise
robustness, outperforming the conventional algorithms, such as CS-BP
\cite{CS-BP2}, in the support detection. In this BHT detection, the
nBP part takes a role to provide the marginal posterior of $X_i$.
Therefore, the advantage of BHT-BP in support detection can be
claimed when the BP convergence is achieved with the sampling rate,
$\frac{M}{N}$, above a certain threshold.

\begin{figure}[!t]
\centering
\includegraphics[width=8.85cm]{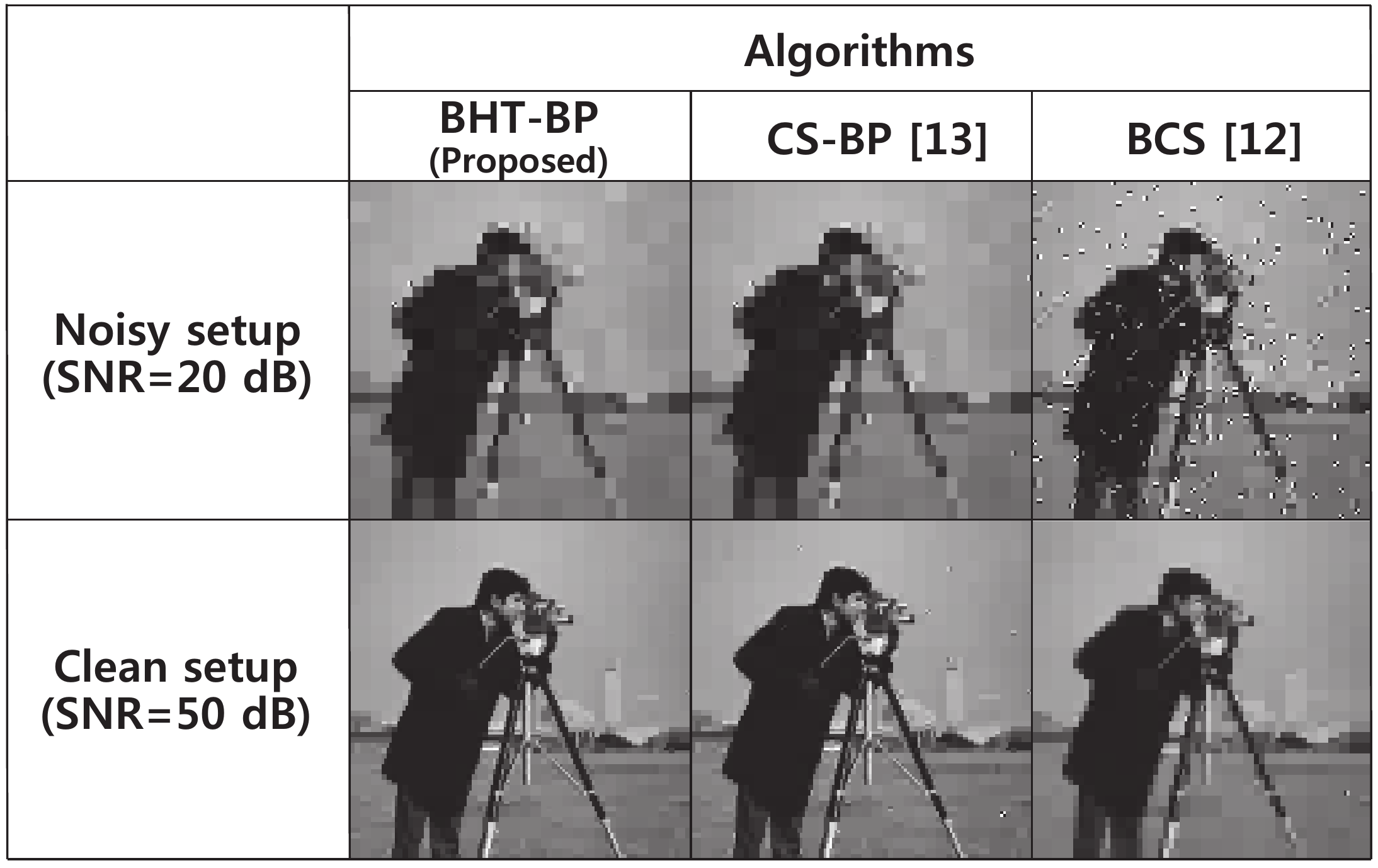}
\caption{An illustrative recovery example of  BHT-BP (the proposed),
CS-BP  \cite{CS-BP2} and BCS \cite{BCS} in the presence of noise.
The original image, known as the Cameraman, of size $N=128^2$, is
transformed via three step discrete wavelet transform. For this
example, we pad zeros for the coefficients having values below $100$
in wavelet domain, and we recover these images from $M/N=0.5$
undersampled measurements. From this example, we  see  that  the
recovered image via BCS includes flicker noise but which is not
shown in that of BHT-BP in the noisy setup. In the clean setup,
BHT-BP more clearly recovers the image than those of CS-BP and BCS.
} \label{fig1}
\end{figure}

Typically, recovery performance of the nBP-based algorithms is
dominated by the message sampling methods. In the case of CS-BP
\cite{CS-BP2}, its performance is corrupted by quantization errors
because CS-BP works with the uniform sampling-based nBP such that
the signal estimate is directly obtained from a sampled posterior.
The joint detection-and-estimation structure of BHT-BP overcomes
this weakpoint of CS-BP, improving  MSE performance. The key behind
the improvement is that  the sampled posterior is only used for the
support detection in BHT-BP. Furthermore, BHT-BP closely approaches
to the oracle performance\footnote{Here, the oracle performance
means the performance of the LMMSE estimator having the knowledge of
the sparse support set of the signal $\mathbf{X}$.}, in high SNR
regime, if the rate $\frac{M}{N}$ are sufficiently maintained for
the signal sparsity $K$. Fig.\ref{fig1} is an illustration intended
to see a motivational evidence of the recovery performance among the
proposed BHT-BP, CS-BP \cite{CS-BP2} and BCS \cite{BCS}.

The importance of the minimum nonzero value $x_{\min}$ of sparse
signals $\mathbf{X}$ in the NSR problems was highlighted by
Wainwright \emph{et al.} in \cite{Wainwright1},\cite{Wainwright2}
 and Fletcher \emph{et al.} in \cite{Fletcher},
where they proved that the perfect support recovery is very
difficult even with arbitrarily large \emph{signal-to-noise ratio}
(SNR) if $x_{min}$ is very small. Following these works, in the
present work, we consider recovery of $\mathbf{X}$ whose minimum
nonzero value is regulated by $x_{min}$. In addition, we propose to
use a signal prior including the parameter $x_{min}$, called
\emph{spike-and-dented slab} prior, investigating how the
performance varies according to the parameter $x_{min}$. We
empirically show in the BHT-BP recovery\footnote{To the best of our
knowledge, we have not seen CS algorithms including $x_{\min}$ as an
input parameter.} that variation of $x_{min}$ is reflected to the
recovery performance in the form of SNR shift. In addition, we
support this statement with a success rate analysis for the BHT
support detection under the identity measurement matrix assumption,
\emph{i.e.}, $\mathbf{\Phi}=\mathbf{I}$.

The recovery complexity of BHT-BP is $\mathcal{O}(N \log N + KM)$
which includes the cost $\mathcal{O}(KM)$ of  the LMMSE estimation
and that of the BHT support detection $\mathcal{O}(N \log N)$. This
is advantageous  compared to that of the $l_1$-norm solvers
$\Omega(N^3)$ \cite{candes2},\cite{Lasso} and BCS $\mathcal{O}(N
K^2)$ \cite{BCS}, being comparable to that of the recent BP-based
algorithms using sparse measurement matrices $\mathcal{O}(N \log
N)$: CS-BP \cite{CS-BP2}  and SuPrEM \cite{SuPrEM}.

\subsection{Organization}

The remainder of the paper is organized as follows. We first provide
basic setup for our work in Section II. In Section III, we discuss
our solution approach to the NSR problem. Section IV
describes a nonparametric implementation of the BHT support detector
and its computational complexity. Section V provides experimental
validation to show performance and several aspects of the proposed algorithm,
compared to the other related algorithms. Finally, we conclude this
paper in Section VI.

\section{Basic Setup}
In this section, we  introduce our signal model, and a factor graphical
model for linear systems used in this work.

\subsection{Signal Model}
Let $\mathbf{x}_0 \in \mathbb{R}^N$
denote a sparse vector which is a deterministic realization of a
random vector $\mathbf{X}$. Here, we assume that the elements of
$\mathbf{X}$ are \emph{i.i.d.}, and each $X_i$ belongs to the
support set  with a sparsity rate $q \in [0,1)$. To indicate the
supportive state of $\mathbf{X}$, we use a state vector $\mathbf{S}
\in \{0,1\}^N$ whose each element $S_i$ is  Bernoulli random with
the rate $q$ as following
\begin{align}\label{eq:eq2-1}
{S_i} = \left\{ \begin{array}{l}
1,\,\,\,\,\,{\rm{if}}\,\, X_i \ne 0 \text{ with } q\\
0,\,\,\,\,\,{\rm{if}}\,\, X_i = 0 \text{ with } 1-q
\end{array} \right..
\end{align}
Then, the signal sparsity,  $K=||\mathbf{S}||_0$, becomes  Binomial
random with $\mathcal{B}(k;N,q)$. In the present work, we consider
the signal $\mathbf{x}_0$ whose minimum nonzero value is regulated
by a parameter $x_{min} \geq 0$. For such signal generation,
\begin{itemize}
\item We first draw a state vector $\mathbf{s}$ by generating $N$
\emph{i.i.d.} Bernoulli numbers of \eqref{eq:eq2-1}.
\item Then,  we
assign zero value  to the signal scalars corresponding to $s_i=0$,
\emph{i.e.},  $x_{0,i}=0$.
\item For the signal scalar corresponding
to $s_i=1$,  a Gaussian number is drawn from
$\mathcal{N}(x;0,\sigma_X^2)$ and assigned to the signal scalar
$x_{0,i}$ if $|x_{0,i}| \geq x_{min}$; otherwise,  the number is
redrawn until a realization with $|x_{0,i}| \geq x_{min}$ occurs.
\end{itemize}

\begin{figure}
\centering
\includegraphics[width=7cm]{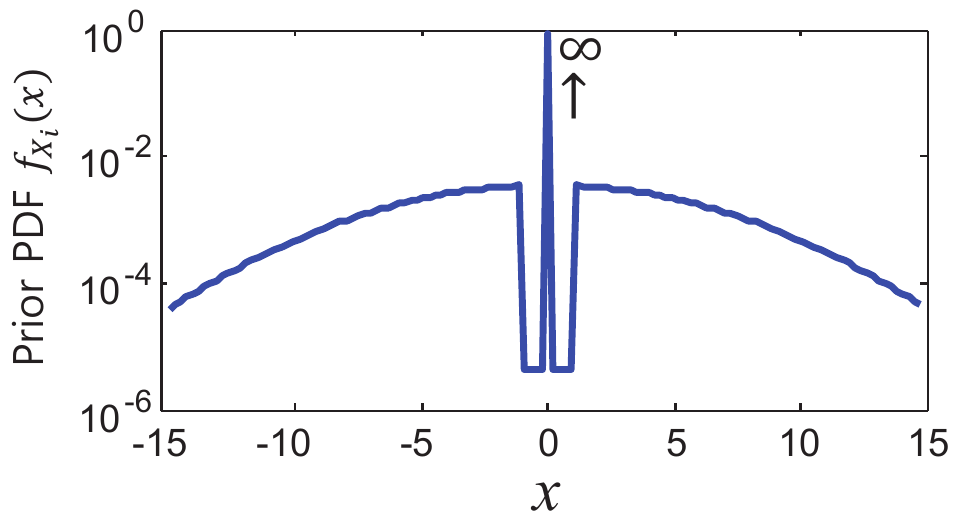}
\vspace{-8pt} \caption{Example of spike-and-dented slab  PDF in
log-scale where the prior is drawn with the parameters, $q=0.05,
\sigma_X=5, x_{\min }=\frac{\sigma_X}{4}, \lambda=10^{-4}$, and
normalized to be $\int_{{X_i}} {{f_{{X_i}}}(x)dx = 1}$.}
\label{fig:prior}
\end{figure}

For such signals with $x_{min}$, we propose to use a
\emph{spike-and-dented slab} prior which is a variant of the
spike-and-slab prior \cite{spikeandslab}. According to
\eqref{eq:eq2-1},  the signal prior of $X_i$ can be described as a
two-state mixture PDF with the state $S_i$, \emph{i.e.},
\begin{align}\label{priorPDF1}
f_{X_i}(x)=(1-q)f_{X_i}(x|S_i=0)+q f_{X_i}(x|S_i=1).
\end{align}
Then, the spike-and-dented slab prior includes the conditional
priors as following
\begin{align}\label{prior}
&{f_{{X_i}}}(x|{S_i} = 0) = \delta(x),\\
&{f_{{X_i}}}(x|{S_i} = 1) \propto \left\{ \begin{gathered}
  \mathcal{N}(x;0,\sigma _X^2),\,\,\,|x| \geq {x_{\min }} \hfill \nonumber \\
  \lambda, \,\,\,\,\,\,\,\,\,\,\,\,\,\,\,\,\,\,\,\,\,\,\,\,\,\,\,\,\,|x| < {x_{\min }} \hfill \\
\end{gathered}  \right.
\end{align}
where $\delta(x)$ is the Dirac delta PDF and $\lambda>0$ is a
near-zero constant. Fig.\ref{fig:prior} shows an example of the
spike-and-dented slab prior where the prior is drawn with the
parameters, $q=0.05, \sigma_X=5, x_{\min }=\frac{\sigma_X}{4},
\lambda=10^{-4}$, and normalized to be $\int_{{X_i}}
{{f_{{X_i}}}(x)dx = 1}$.



\begin{figure}[!t]
\centering
\includegraphics[width=9cm]{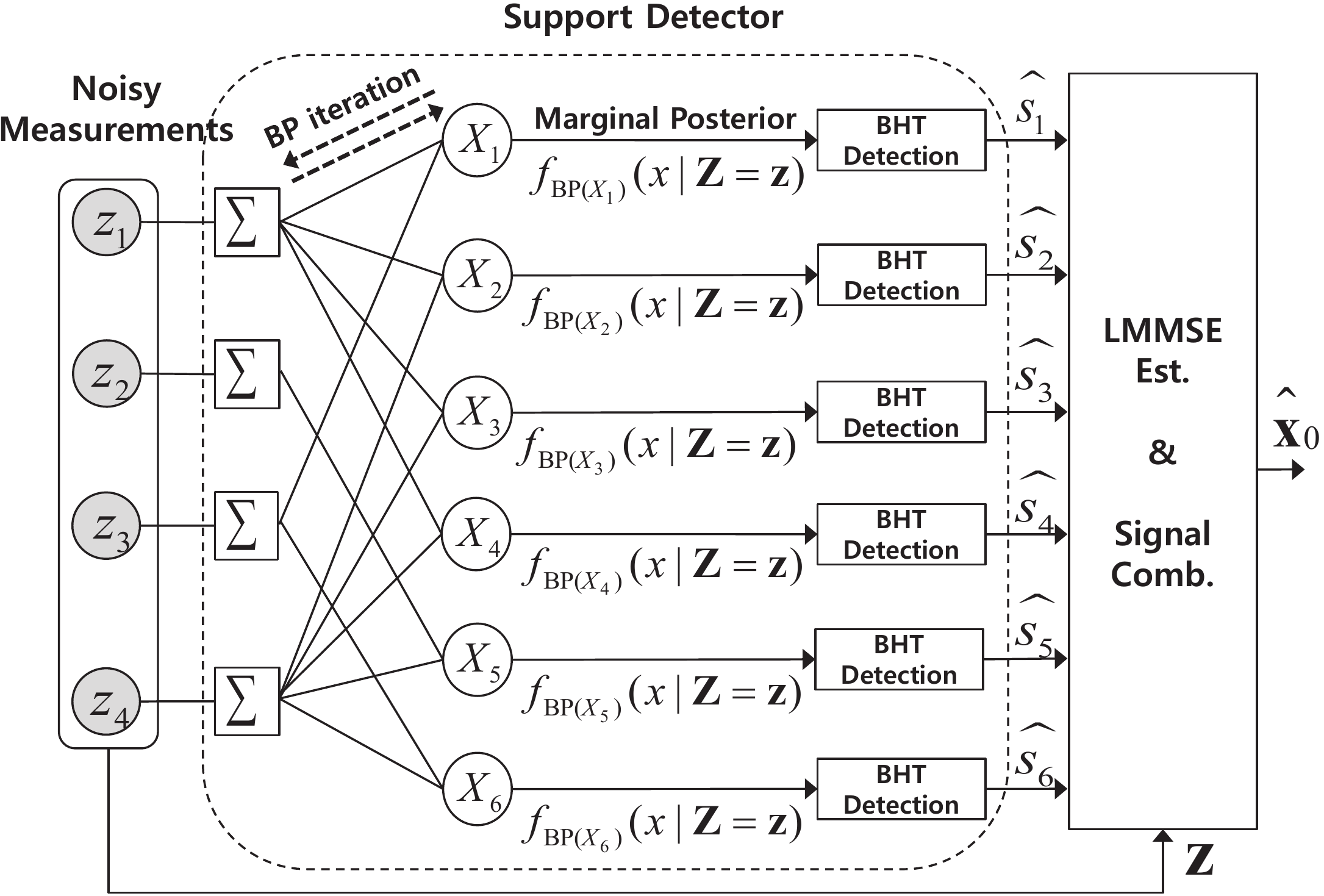}
\caption{Diagrammatic representation of the proposed algorithm (when
$N=6,M=4,L=2$) where the inputs of the proposed algorithm is the
measurement $\mathbf{z}=[z_1,z_2,z_3,z_4]$ and the output is a
signal estimate $\widehat{\mathbf{x}}_0$. The proposed algorithm
first detects the signal support $\widehat{\mathbf{s}}=[\widehat
{s}_1,...,\widehat {s}_6]$ from the measurements $\mathbf{z}$ at
hand, and then applies linear MMSE estimation to find the signal
estimate $\widehat{\mathbf{x}}_0$ given the detected support
$\widehat{\mathbf{s}}$.} \label{fig:Fig2-1}
\end{figure}

The goal of the proposed algorithm is to recover the signal vector
$\mathbf{x}_0$ from a noisy measurement vector
\begin{eqnarray}\label{eq:eq2-2}
\mathbf{z}=\mathbf{\Phi}\mathbf{x}_0 + \mathbf{w} \in \mathbb{R}^M,
\end{eqnarray}
given a fat measurement matrix $\mathbf{\Phi} \in \{0,1,-1\}^{M \times
N}$ $(M<N)$, where the vector $\mathbf{w} \in \mathbb{R}^M$ is a
realization of a Gaussian random vector $\mathbf{W} \sim
{\mathcal{N}(\mathbf{w};0,\sigma_W^2\mathbf{I})}$; therefore, the
vector $\mathbf{z} \in \mathbb{R}^M$ is drawn from a mean shifted
Gaussian random vector conditioned on $\mathbf{X}=\mathbf{x}_0$, \emph{i.e.}, $\mathbf{Z} \sim
{\mathcal{N}(\mathbf{z};\mathbf{\Phi}\mathbf{x}_0,\sigma_W^2\mathbf{I})}$.
For the measurement matrix $\mathbf{\Phi}$, we consider  an LDPC-like
sparse matrix   which has very low matrix sparsity (typically less
than 1$\%$ matrix sparsity) and the tree-structured property
\cite{Mackey},\cite{Richardson}. We regulate the matrix sparsity by the fixed column
weight $L$ such that $\mathbb{E}[\left\| {{\phi _{\text{column}}}}
\right\|_2^2]=L$. This regulation enables the matrix $\mathbf{\Phi}$
to span the measurement space with column vectors having equal
energy.

\subsection{Factor Graphical Modeling of Linear Systems}
Factor graphs effectively represent such sparse linear systems in
\eqref{eq:eq2-2} \cite{factor}. Let $\mathcal{V}:=\{1,...,N\}$
denote a set of variable nodes   corresponding to the signal
elements, $\mathbf{x}_0=[x_{0,1},...,x_{0,N}]$, and
$\mathcal{C}:=\{1,...,M\}$ denote a set of factor nodes
corresponding to the measurement elements,
$\mathbf{z}=[z_1,...,z_M]$. In addition, we define a set of edges
connecting $\mathcal{V}$ and $\mathcal{C}$ as $\mathcal{E} := \{
(j,i) \in \mathcal{C} \times \mathcal{V}\,|\,\,\phi _{ji} = 1\}$
where $\phi _{ji}$ is the $(j,i)$-th element of $\mathbf{\Phi }$.
Then, a factor graph $\mathcal{G}=(\mathcal{V,C,E})$ fully describes
the neighboring relation in the sparse linear system. For
convenience, we define the neighbor set of $\mathcal{V}$ and
$\mathcal{C}$ as $N_{\mathcal{V}} (i) := \{ j \in
\mathcal{C}\,|(j,i) \in {\mathcal{E}}\}$ and $N_\mathcal{C} (j) :=
\{ i \in \mathcal{V}\,|(j,i) \in {\mathcal{E}}\}$, respectively.
Note that the column weight of  the matrix $\mathbf{\Phi}$ is
expressed as $L=\left|N_{\mathcal{V}} (i)\right|$ in this graph
model.

\section{Solution Approach of Proposed Algorithm}

The proposed algorithm, BHT-BP, has a joint detection-and-estimation
structure where we first detect the sparse support by a combination
of BP and BHT, then estimating nonzeros in the detected support by
an LMMSE estimator, as shown in Fig.\ref{fig:Fig2-1}. In this
section, we provide our solution approach to the support detection
and the nonzero estimation under the joint structure.

\subsection { Support Detection using Bayesian Hypothesis Testing}
\subsubsection {Support detection in BHT-BP}
The support detection  problems can be decomposed to a sequence of
binary state detection problems  given the marginal posterior
$f_{{X_i}}(x|{\bf{Z}} = {\bf{z}})$ of each signal scalar. Our state
detection problem is to choose one between the two hypotheses:
\begin{align}
\begin{array}{l}
\mathcal{H}_0:S_i=0\,\,\,\text{and}\,\,\,\mathcal{H}_1:S_i=1,\nonumber
\end{array}
\end{align}
given the measurements $\mathbf{z}$. Our methodlogy to this problem
is related to \emph{Bayesian composite hypothesis testing} [32, p.
198]. In contrast to the simple hypothesis test where the PDFs under
both hypothesis are perfectly specified, the composite hypothesis
test must  consider associated random variables. In our problem, the
associated random variable is $X_i$. Then, the binary state detector
decides $\mathcal{H}_1$ if
\begin{align}\label{BHT1}
\frac{ {   {f_{\mathbf{Z}}}({\mathbf{z}}|{\mathcal{H}_1})   }}{
{f_{\mathbf{Z}}}({\mathbf{z}}|{\mathcal{H}_0}) }
=\frac{{\int {
{{f_{\mathbf{Z}}}({\mathbf{z}}|{\mathcal{H}_1},{X_i} = x)}
{f_{{X_i}}}( x|{\mathcal{H}_1}) dx} }}{{\int {
{{f_{\mathbf{Z}}}({\mathbf{z}}|{\mathcal{H}_0},{X_i} = x)}
{f_{{X_i}}}( x|{\mathcal{H}_0})dx} }} > \gamma,
\end{align}
where $\gamma$ is a threshold for the test.
The PDF
${{f_{\mathbf{z}}}({\mathbf{z}}|{\mathcal{H}_{s_i}},{X_i} = x)}$ is
simplified to ${{f_{\mathbf{Z}}}({\mathbf{z}}|{X_i} = x)}$ since the
hypothesis $\mathcal{H}_{s_i}$ and the measurements $\mathbf{Z}$ are
conditionally independent given $X_i$. Therefore, finally, the
binary hypothesis test in \eqref{BHT1} can be rewritten as
\begin{align}\label{eq:eq2-7}
T_{\text{\tiny BHTBP}}(\mathbf{z}):=\frac{{\int
{\frac{{{f_{{X_i}}}(x|S_i=1)}}{{{f_{{X_i}}}(x)}}{f_{{X_i}}}(x|{\mathbf{Z}}
= {\mathbf{z}})dx} }}{{\int
{\frac{{{f_{{X_i}}}(x|S_i=0)}}{{{f_{{X_i}}}(x)}}{f_{{X_i}}}(x|{\mathbf{Z}}
= {\mathbf{z}})dx} }} \mathop {\mathop \gtrless
\limits_{{\mathcal{H}_0}} }\limits^{{\mathcal{H}_1}}\gamma
\end{align}
where the Bayesian rule is applied to
${{f_{\mathbf{Z}}}({\mathbf{z}}|{X_i} =
x)}=\frac{{{f_{{X_i}}}(x|{\mathbf{Z}} =
{\mathbf{z}}){f_{{\mathbf{Z}}}}(\mathbf{z})}}{{{f_{{X_i}}}(x)}}$,
and obviously
${f_{{X_i}}}(x|{\mathcal{H}_{s_i}})={f_{{X_i}}}(x|S_i=s_i)$ holds
from the prior knowledge of \eqref{priorPDF1}.

\begin{figure}[!t]
\centering
\includegraphics[width=9cm]{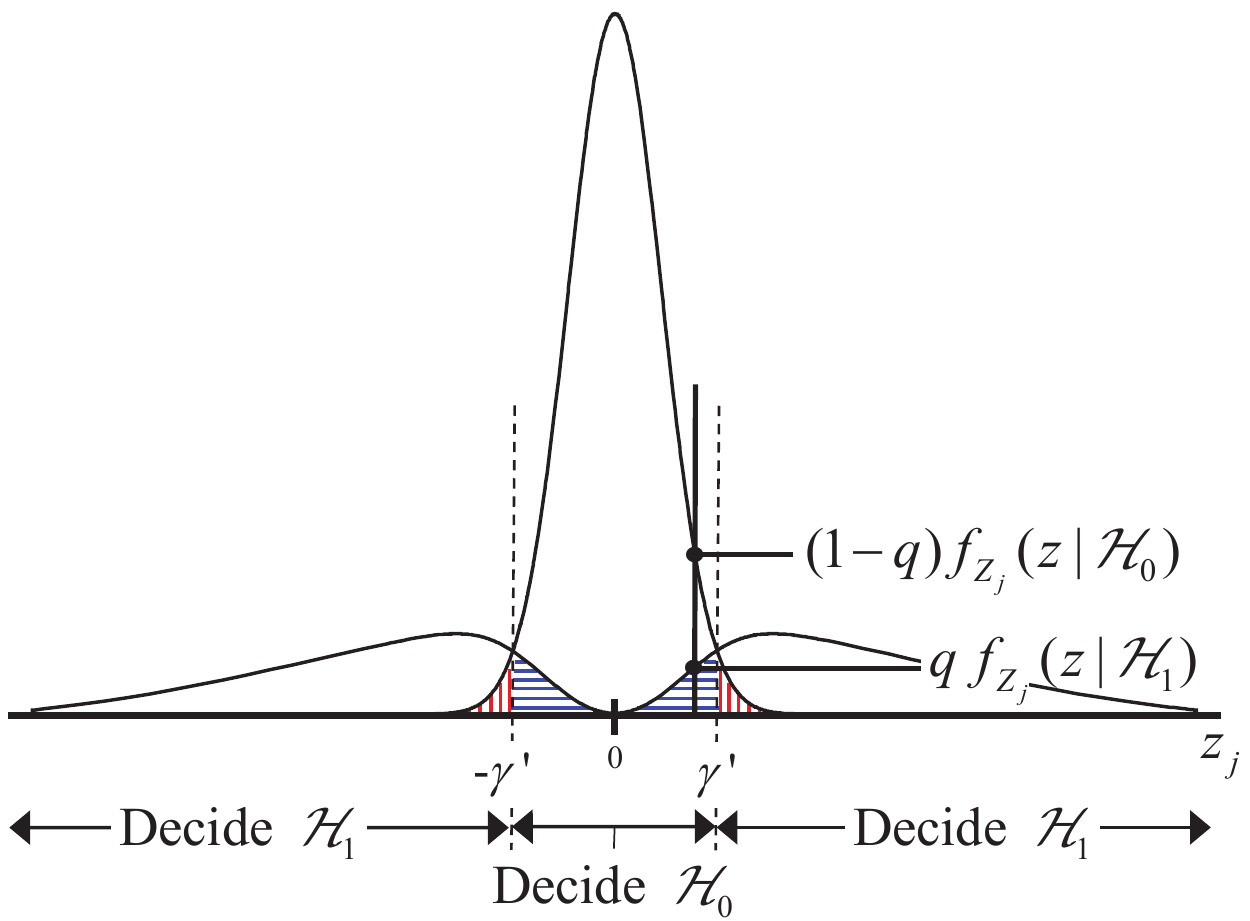}
\caption{Fig.\ref{fig3-5} illustrates  the scalar state detection by
BHT under an assumption of $\mathbf{\Phi}=\mathbf{I}$. Under this
assumption, ``the hypothesis test given a vector $\mathbf{z}$" is
simplified to ``the test given a  scalar $z_j$", described in
\eqref{scalar_test}, where the threshold $\gamma'$ is derived from
the equality condition of \eqref{eqcondition}. In the figure, the
horizontal-lined region (blue) represents $ \Pr \{ {\widehat s_i}
\ne {s_i}|{\mathcal{H}_1}\}$ and the vertical-lined (red) region
does $\Pr \{ {\widehat s_i} \ne {s_i}|{\mathcal{H}_0}\}$.}
\label{fig3-5}
\end{figure}

In some detection problem under Bayesian paradigm, one can
reasonably assign prior probabilities to the hypotheses. In the
present work, we assign the sparsity rate $q$ to the hypotheses,
\emph{i.e.}, $\Pr\{\mathcal{H}_0\}=1-q$ and
$\Pr\{\mathcal{H}_1\}=q$. Then, we can define the state error rate
(SER) of the scalar state detection \eqref{eq:eq2-7} [32, p. 78]
\begin{align}\label{SER}
{P_{\text{SER}}} := \Pr \{ {\widehat s_i} \ne
{s_i}|{\mathcal{H}_0}\} (1-q)  + \Pr \{ {\widehat s_i} \ne
{s_i}|{\mathcal{H}_1}\} q.
\end{align}
It is well known that  the threshold $\gamma$ of \eqref{eq:eq2-7}
can be optimized under \emph{the criterion of the minimum  detection
error probability} with the SER expression \eqref{SER}. By the
criterion, we assign the threshold to $\gamma=\gamma^*:=
\frac{1-q}{q}$. We omit the derivation for this threshold
optimization here, referring interested readers to  [32, p. 90]. We
call this binary hypothesis test \eqref{eq:eq2-7} with the threshold
$\gamma^*$ as \emph{Bayesian hypothesis test} (BHT) detection.  The
proposed algorithm generates a detected support $\widehat
{\mathbf{s}} \in \{0,1\}^N$ according to the results of a sequence
of BHTs. Therefore,  given a marginal posterior of each $X_i$,
BHT-BP can robustly detect the signal support even when the
measurements are noisy.

Fig.\ref{fig3-5}   illustrates the scalar state detection of BHT-BP
when the matrix is $\mathbf{\Phi}=\mathbf{I}$ such that the
measurement channel can be decoupled to $N$ scalar Gaussian
channels, \emph{i.e.}, $Z_j=X_i+W_j,\, (i=j)$. Under this
assumption, ``the hypothesis test given a vector $\mathbf{z}$" can
be scalarwise to ``the test given a  scalar $z_j$", being simplified
\begin{align}\label{scalar_test}
\forall j \in\mathcal{C}: |z_j|  \mathop {\mathop \gtrless
\limits_{{\mathcal{H}_0}} }\limits^{{\mathcal{H}_1}} \gamma'
\end{align}
where the threshold $\gamma'$ is derived from the equality condition
with the two scalar likelihood and the threshold
$\gamma^*=\frac{1-q}{q}$,
\begin{align}\label{eqcondition}
\frac{ { {f_{Z_j}}(z|{\mathcal{H}_1}) }}{
{f_{Z_j}}(z|{\mathcal{H}_0}) } =\gamma^*.
\end{align} Hence, the threshold $\gamma'$ is a function of
$\sigma_X$, $\sigma_W$, $x_{\min}$, and $q$ (see Appendix II). With
this threshold $\gamma'$, we can find the conditional SER, $\Pr \{
{\widehat s_i} \ne {s_i}|{\mathcal{H}_{s_i}}\}$, for the case
$\mathbf{\Phi}=\mathbf{I}$. In Fig.\ref{fig3-5}, the
horizontal-lined region (blue) represents $ \Pr \{ {\widehat s_i}
\ne {s_i}|{\mathcal{H}_1}\}$ and the vertical-lined (red) region
does $\Pr \{ {\widehat s_i} \ne {s_i}|{\mathcal{H}_0}\}$. The
corresponding SER analysis will be provided in  Appendix II.
Although Fig.\ref{fig3-5} does not show typical behavior of the BHT
detection given a vector measurement $\mathbf{z}$, the figure helps
intuitive understanding of the BHT detection.

In addition, it is noteworthy in Fig.\ref{fig3-5} that the shape of
${f_{Z_j}}(z|{\mathcal{H}_1})$ is dented near $z_j=0$. This is
caused by the use of  the spike-and-dented slab prior, given in
\eqref{prior}, where the dented part varies with the parameter
$x_{\min}$.

\bigskip
\subsubsection {Support detection of CS-BP}
Support detection is not performed in practical recovery of CS-BP,
but we describe it here for a comparison purpose. CS-BP estimates
the sparse solution $\widehat{\mathbf{x}}_0$ directly from a BP
approximation of the  signal posterior, through MAP or MMSE
estimation. Let us consider CS-BP using the MAP estimation. Then,
given the marginal posterior $f_{{X_i}}(x|{\bf{Z}} = {\bf{z}})$, the
scalar state detection of CS-BP is equivalent to choose one of the
two peaks at $x=0$ and $x=\widehat x_{\text{MAP},i}: = \arg \mathop
{\max }\limits_{x} f_{{X_i}}(x|{\bf{Z}} = {\bf{z}})$. Namely, the
binary state detector of CS-BP decides $\mathcal{H}_1$ if
\begin{align}\label{CSBP_suppdet}
T_{\text{\tiny CSBP}}(\mathbf{z})&:=\frac{{\Pr \{ \widehat x_{\text{MAP},i}-\Delta x < {X_i} \leq \widehat x_{\text{MAP},i}+\Delta x|\mathbf{Z}=\mathbf{z} \} }}{{\Pr \{ 0-\Delta x < {X_i} \leq 0+\Delta x|\mathbf{Z}=\mathbf{z}\} }}\nonumber\\
&= \frac{{\int_{\widehat x_{\text{MAP},i}-\Delta x}^{\widehat x_{\text{MAP},i}+\Delta x} { f_{{X_i}}(x|{\bf{Z}} = {\bf{z}}) dx} }} {{ \int_{0-\Delta x}^{0+\Delta x} f_{{X_i}}(x|{\bf{Z}} = {\bf{z}}) dx} } > 1,
\end{align}
where $\Delta x$ is a small quantity that we eventually let approach
to 0. When $\widehat x_{\text{MAP},i}=0$, the test cost becomes one;
then, the detector immediately decides $\mathcal{H}_0$. Hence, in
CS-BP, the detected support $\mathbf{\widehat s}$ is just a
by-product of the signal estimate
$\widehat{\mathbf{x}}_{\text{MAP}}$, which is not robust support
detection against additive measurement noise.

\subsection {Conditions for BP Convergence}
In the proposed algorithm,  marginal posterior of each $X_i$,
$f_{X_i}(x|\mathbf{Z}=\mathbf{z})$, for the BHT detection  is
computed by BP. It was known that BP efficiently computes such
marginal posteriors, achieving its convergence if the conditions in
Note 1 are satisfied \cite{Bishop}. Given the BP convergence, each
approximate marginal posterior converges to a PDF peaked at an
unique value $\widehat x_{i}$ during the iteration. In noiseless
setup,
the unique value is exactly the true value, \emph{i.e.}, $\widehat x_{i}=x_{0,i}$.\\
\\
\emph{ \bf{Note 1 (Conditions for BP convergence):} }
\begin{itemize}
\item The  factor graph, which corresponds to the relation between $\mathbf{X}$ and $\mathbf{Z}$, has a tree-structure.
\item Sufficiently large number of iterations $l$ is maintained such that BP-messages have been
propagated along every link of the tree, and a variable node has
received messages from all the other variables nodes.
\end{itemize}
Although the second condition in Note 1 is practically demanding, it
has been reported  that BP provides a good approximation of marginal
posteriors even with factor graphs including cycles, which is called
loopy BP \cite{Mackey},\cite{Freg},\cite{loopyBP}.

A related argument for BP was stated by Guo and Wang in the context
of the multiuser detection problem of CDMA systems, where the
problem is actually equivalent to solve a linear system
\cite{Guo05},\cite{Guo08}. In the works, Guo and Wang showed that
the marginal posterior computed by BP is almost exact in a large
linear system ($M,N \to \infty$) if the factor graph corresponding
to the  matrix $\mathbf{\Phi}$ is asymptotically cycle-free and  the
sampling rate $\frac{M}{N}$ is above a certain threshold\footnote{In
\cite{Guo05},\cite{Guo08}, the authors considered the sampling rate
$\frac{M}{N}$ above one.}. Namely, Guo and Wang showed that
\begin{align}\label{eq:eq2-4}
\mathop {\lim }\limits_{l \to \infty } \mathop {\lim \sup
}\limits_{M,N \to \infty } \left|
{f_{\text{BP}({X_i})}^{(l)}(x|{\bf{Z}} = {\bf{z}}) -
f_{{X_i}}^{}(x|{\bf{Z}} = {\bf{z}})} \right| = 0,
\end{align}
where $f_{\text{BP}({X_i})}^{(l)}(x|{\bf{Z}} = {\bf{z}})$ is an
approximate marginal posterior of each $X_i$ by  $l$ iterations of
BP.

According to the literature, in the linear system with the LDPC-like
matrix $\mathbf{\Phi}$, the sampling rate $\frac{M}{N}$ is the only
obstacle  for the BP convergence. The asymptotic condition used in
\eqref{eq:eq2-4} is not always necessary if the tree-structured
property is guaranteed for the matrix $\mathbf{\Phi}$ because the
main reason to use the asymptotic condition in the works of
\cite{Guo05},\cite{Guo08} is to make the system  ``asymptotically
cycle-free'', which is equivalent to having an ``asymptotically
tree-structured'' matrix $\mathbf{\Phi}$\footnote{If the
 graph corresponding to the matrix $\mathbf{\Phi}$ has at least one cycle, the
BP convergence cannot be rigorously guaranteed.}. Thus, we claim the
advantage of BHT-BP over CS-BP in support detection with a certain
threshold of the rate $\frac{M}{N}$. Given $\frac{M}{N}$ below the
threshold, the BP convergence is not achieved such that the
likelihood $f_{\mathbf{Z}}(\mathbf{z}|\mathcal{H}_{s_i})$ is not
properly calculated for the BHT detection. We will empirically find
the threshold using information entropy of the approximate marginal
posterior, $f_{\text{BP}({X_i})}^{(l)}(x|{\bf{Z}} = {\bf{z}})$, in
Section V-A. Although we do not provide an analytical threshold of
$\frac{M}{N}$ for the BP convergence in this paper, simulation
results with $\frac{M}{N}$ above the empirical threshold are quite
favorable, as shown in Section V-B and -C.

\subsection{LMMSE Estimation of Nonzero Values} Given the
support information by the BHT detection, the rest of the work is
reduced to the nonzero estimation problem, represented as
\begin{align}\label{eq:eq2-9}
&{\widehat {\bf{x}}_{0}} = \mathbb{E}\left[ {{{\bf{X}}}|{\bf{S}} =
{\widehat{\bf{s}}},{\bf{Z}}={\bf{z}}} \right],
\end{align}
and it can  straightforwardly solved by combining the nonzero
position by $\widehat{ \mathbf{s}}$ and the nonzero values given by
the LMMSE estimate [33, p. 364]
\begin{align}\label{eq:eq2-10}
{\widehat {\bf{x}}_{0,\widehat{\bf{s}} }} = {\left(
{\frac{1}{{\sigma_{X}^2}}{\bf{I}} + \frac{1}{{\sigma _W^2}}{\bf{\Phi
}}_{ \widehat{\bf{s}} }^T{\bf{\Phi }}_{
 \widehat{\bf{s}}}} \right)^{ - 1}}
\frac{1}{{\sigma _W^2}}{\bf{\Phi }}_{ \widehat{\bf{s}}}^T\mathbf{z},
\end{align}
where $\mathbf{\Phi}_{\widehat{\mathbf{s}}} \in \{0,1\}^{M \times
K}$ denotes a submatrix of $\mathbf{\Phi}$ that contains only the
columns corresponding to the detected support
$\widehat{\mathbf{s}}$, $\sigma_X^2$ are the  variance of an nonzero
scalar $X_i$.

The estimate ${\widehat {\bf{x}}_{0}}$ from the proposed joint
detection-and-estimation structure is not optimal. As we have seen,
our support detector \eqref{eq:eq2-7} is based on the criterion of
minimum detection error probability. Even with this detector,
however, we cannot guarantee the estimation optimality since the
LMMSE estimator of \eqref{eq:eq2-9} is not designed from the cost
function involving the detection part
\cite{DE_info},\cite{Moustaki}. Nevertheless, worth mentioning here
is that the proposed joint structure has advantages as given in Note
2.\\
\\
\emph{ \bf{Note 2 (Claims from the joint detection-and-estimation
structure):} }
\begin{itemize}
\item Removing  the MSE degradation
caused by the uniform sampling-based nBP.
\item Achieving the oracle performance in the high
SNR regime with the sufficiently high rate $\frac{M}{N}$ for the BP
convergence.
\end{itemize}
We will empirically validate this claim in Section V-C.

\section{Nonparametric Implementation of BHT Support Detector}
This section describes a nonparametric implementation of the
proposed support detector consisting of BP and the BHT detection. We
discuss our nonparametric approach of the BP part first, and then
explain the BHT detection part. This BHT support detection is
summarized in Algorithm \ref{algo1}.

\subsection{Nonparametric  BP  using Uniform Sampling}
In the BHT support detector, the BP part  provides the marginal
posterior of $X_i$ for the hypothesis test in \eqref{eq:eq2-7}.
Since the signal $\mathbf{x}_0$ is real valued, each BP-message
takes the form of a PDF, and the BP-iteration becomes a
density-message-passing process.  To implement  the
density-message-passing, we take the nBP approach
\cite{non_para_BP}-\cite{Noorshams}. Many nBP algorithms have been
proposed according to several message sampling methods such as
discarding samples having low probability density \cite{coughlan},
adaptive sampling \cite{Isard}, Gibbs sampling \cite{non_para_BP},
rejection sampling \cite{Noorshams}, or importance sampling
\cite{nBP_sensor}.

Our nBP approach is to use an uniform sampling for the
message representation where we set the sampling step $T_s$ on the
basis of \emph{the three sigma-rule} \cite{sigmarule} such that
\begin{eqnarray} \label{eq:eq3-14}
T_s=\frac{2 \cdot 3 \sigma_{X}}{N_d}
\end{eqnarray}
where $N_d$ is the number of samples to store a BP-message.
Then, we define the uniform sampling of a density-message $f(x)$ as
\begin{align}\label{eq:eq3-15}
{\rm{Samp}}\left\{ {f(x)};T_s \right\} &:= f(m{T_s} -
3{\sigma_{X}})\nonumber\\
&=f[m], \,\,{\rm{ for }}\,\,m \in \{0,1,...,{N_d}-1\},
\end{align}
where $\rm{Samp}\left\{ \cdot;T_s\right\}$ denotes the uniform
sampling function with the step size $T_s$. Hence, the sampled
message $ f[m]$ can be treated as a vector with size $N_d$ by
omitting the index $m$. The main strength of the uniform sampling-based nBP is
adaptivity to various signal prior PDFs.  In addition,  we note that
the calculation of uniformly sampled messages can be accelerated
using the \emph{Fast fourier transform} (FFT).

Consider the factor graph $\mathcal{G}=(\mathcal{V,C,E})$ depicted
in the support detection part of Fig.\ref{fig:Fig2-1} where a signal
element $X_i$ corresponds to a variable node $i\in\mathcal{V}$ and a
measurement element $Z_j$ corresponds to a factor node
$j\in\mathcal{C}$. At every iteration, messages are first passed
from each variable node $i \in \mathcal{V}$ to its neighboring
factor nodes $N_{\mathcal{V}}(i)$; each factor nodes $j \in
\mathcal{C}$ then calculates messages to pass back to the
neighboring variable nodes $N_{\mathcal{C}}(j)$ based on the
previously received messages. These factor-to-variable (FtV)
messages include \emph{extrinsic} information of $X_i$, and will
then be employed for the computation of updated variable-to-factor
(VtF) messages in the next iteration. (For the detail, see the paper
\cite{factor}).

Let $\mathbf{a}_{i \rightarrow j}^{(l)} \in [0,1)^{N_d}$ denote a
sampled VtF message at the $l$-th iteration in the vector form,
given as
\begin{eqnarray}\label{eq:eq3-16}
\mathbf{a}_{i \rightarrow j}^{(l)}=\eta\left[ {{\mathbf{p}_{X_i}}
\times \prod\limits_{k \in N_{\mathcal{V} }(i)\backslash\{j\}}
{{\bf{b}}_{k \rightarrow i}^{(l-1)} } } \right]
\,\,\forall(j,i)\in\mathcal{E},
\end{eqnarray}
where all product operations are elementwise, the vector
$\mathbf{p}_{X_i} \in [0,1)^{N_d}$ denotes the sampled signal prior,
\emph{i.e.}, $\mathbf{p}_{X_i} :=\text{Samp}\{f_{X_i}(x),T_s \}$,
and $\eta[\cdot]$ is a normalization function to make $
\sum{\mathbf{a}_{i\rightarrow j}^{(l)} }=1$. The sampled FtV message
at the $l$-th iteration, $\mathbf{b}_{j \rightarrow i}^{(l)} \in
[0,1)^{N_d}$, is defined as
\begin{align}\label{eq:eq3-17}
\mathbf{b}_{j \rightarrow i}^{(l)} = \mathbf{p}_{Z_j|\mathbf{X}}
 \otimes \left(\bigotimes
\limits_{k \in N_{\mathcal{C}}(j)\backslash \{i\} } {\mathbf{a}_{k
\rightarrow j}^{(l)}} \right)\,\,\forall(j,i)\in\mathcal{E},
\end{align}
where $\otimes$ is the operator for the linear convolution of
vectors,  and the vector $\mathbf{p}_{Z_j|\mathbf{X}} \in
[0,1)^{N_d}$ is the sampled measurement PDF, \emph{i.e.},
$\mathbf{p}_{Z_j|\mathbf{X}}
:=\text{Samp}\{\mathcal{N}(z_j;(\mathbf{\Phi}\mathbf{X})_j,\sigma_W^2),T_s
\}$.

The convolution operations in \eqref{eq:eq3-17} can be efficiently
computed by using FFT. Accordingly, we can rewrite the FtV message
calculation as
\begin{align}\label{eq:eq3-18}
{\bf{b}}_{j \to i}^{(l)}= \mathcal{F}^{ - 1}  \left\{
\mathcal{F}\mathbf{p}_{Z_j|\mathbf{X}}\times
\left(\prod \limits_{k \in N_{\mathcal{C} }(j)\backslash\{i\} }
{\mathcal{F}\mathbf{a}_{k \rightarrow j}^{(l)}} \right) \right\}
\end{align}
where $\mathcal{F}$ denotes the FFT operation. Therefore, for
efficient use of FFT, the sampling step $T_s$ should be
appropriately chosen such that $N_d$ is power of two. In fact, the
use of  FFT brings a small calculation gap since the FFT-based
calculation performs a circular convolution.  However, this gap can
be ignored, especially when the messages take the form of
bell-shaped PDFs such as Gaussian PDFs.

The sampled approximation of the marginal posterior of each $X_i$,
\emph{i.e.}, $\mathbf{p}_{X_i|\mathbf{Z}}^{(l)}
:=\text{Samp}\{f_{\text{BP}(X_i)}^{(l)}(x|\mathbf{Z}=\mathbf{z}),T_s
\}$, is produced by using the FtV message \eqref{eq:eq3-17} for
every $i\in\mathcal{V}$. Namely,
\begin{align}\label{eq:eq3-19}
\mathbf{p}_{X_i|\mathbf{Z}}^{(l)}=\eta\left[ {{\mathbf{p}_{X_i}}
\times \prod\limits_{k \in N_{\mathcal{V} }(i)} {{\bf{b}}_{k
\rightarrow i}^{(l-1)} } } \right]\,\,\forall i\in\mathcal{V},
\end{align}
To terminate the BP loop, we test the condition at every iteration,
which is given as
\begin{align}\label{eq:eq3-19-2}
\frac{1}{N}\sum\limits_{i = 1}^N {\frac{{\|
{{\bf{p}}_{{X_i}|{\bf{Z}}}^{(l)} - {\bf{p}}_{{X_i}|{\bf{Z}}}^{(l -
1)}} \|_2^2}}{{\| {{\bf{p}}_{{X_i}|{\bf{Z}}}^{(l)}} \|_2^2}}} \leq
\varepsilon
\end{align}
where $\varepsilon >0$ is a constant for the termination condition.
If the condition given in \eqref{eq:eq3-19-2} is satisfied, the BP
loop will be terminated.  After the BP termination, we can simply
express the marginal posterior of $X_i$ by dropping out the
iteration index $l$, \emph{i.e.}, ${\bf{p}}_{{X_i}|{\bf{Z}}}$.

\begin{algorithm}[!t]
\caption{BHT support detection}\label{algo1}
\begin{algorithmic}[0]
\Require Noisy measurements $\mathbf{z}$, measurement matrix
$\mathbf{\Phi}$, sparsity rate $q$, sampled  prior PDF
$\mathbf{p}_X$, sampled measurement PDF $\mathbf{p}_{Z_j|\mathbf{X}}$, The
number of samples $N_d$, Termination condition $\varepsilon$.

\Ensure Reconstructed signal $\widehat{\mathbf{x}}_{0}$, Detected
support vector $\widehat{\mathbf{s}}$.
\\
\State 1) Belief propagation: \State set $\mathbf{b}_{j \rightarrow
i}^{(l=0)}=\mathbf{1}\text{ for all } (j,i) \in \mathcal{E}$
\While{$\frac{1}{N}\sum\limits_{i = 1}^N {\frac{{\left\|
{{\bf{p}}_{{X_i}|{\bf{Z}}}^{(l)} - {\bf{p}}_{{X_i}|{\bf{Z}}}^{(l -
1)}} \right\|_2^2}}{{\left\| {{\bf{p}}_{{X_i}|{\bf{Z}}}^{(l)}}
\right\|_2^2}}}  > \varepsilon$}
\\
$\forall (j,i) \in \mathcal{E}$: \State set $\mathbf{a}_{i
\rightarrow j}^{(l)}=\eta\left[ {{\mathbf{p}_{X_i}} \times
\prod\limits_{k \in N_{\mathcal{V} }(i)\backslash\{j\}} {{\bf{b}}_{k
\rightarrow i}^{(l-1)} } } \right]$ \State set $\mathbf{b}_{j
\rightarrow i}^{(l)} = \mathbf{p}_{Z_j|\mathbf{X}}
 \otimes \left(\bigotimes
\limits_{k \in N_{\mathcal{C}}(j)\backslash \{i\} } {\mathbf{a}_{k
\rightarrow j}^{(l)}} \right)$
\\
$\forall i \in \mathcal{V}$: \State set
$\mathbf{p}_{X_i|\mathbf{Z}}^{(l)}=\eta \left[ {{\bf{a}}_{i \to
j^*}^{(l)} \times {\bf{b}}_{j^* \to i}^{(l - 1)}} \right]$

\EndWhile
\\
\State 2) BHT detection:
\\
$\forall  i \in \mathcal{V}$:\\
        \If {$\log \frac{{\sum {{{\bf{r}}_1} \times {{\bf{p}}_{{X_i}|{\bf{Z}}}}}
}}{{\sum\limits_{} {{{\bf{r}}_0} \times {{\bf{p}}_{{X_i}|{\bf{Z}}}}}
}}\mathop > \log \frac{{1 - q}}{q}$} set $\widehat{s}_i =1$
        \Else  {} set $\widehat{s}_i =0$
        \EndIf
\end{algorithmic}
\end{algorithm}

\begin{table*}
\renewcommand{\arraystretch}{1.3}
\caption{List of algorithms in the performance validation}
\label{table1} \normalsize
 \centering
\begin{tabular}{||c||c|c|c|c||}
\hline
Algorithms & Complexity        & Type of $\Phi$              & Type of Prior PDFs  &    Utilized Techniques \\
\hline \hline
BHT-BP (Proposed)   & $\mathcal{O}(N\log N+KM)$        & LDPC-like                & Spike-and-dented slab   &  nBP    \\
 \hline
CS-BP \cite{CS-BP2}  & $\mathcal{O}(N\log N)$                  & LDPC-like          &  Spike-and-dented slab    &  nBP  \\
\hline
SuPrEM  \cite{SuPrEM}  & $\mathcal{O}(N\log N)$               & LDF                & Two-layer Gaussian  with Jeffery  & EM, pBP\\
\hline
BCS   \cite{BCS}  & $\mathcal{O}(NK^2)$                   & LDPC-like           & Two-layer Gaussian  with Gamma  & EM   \\
\hline $l_1$-DS \cite{candes2} & $\Omega(N^3)$           & Std.
Gaussian & - & CVX
opt.  \\
\hline
\end{tabular}
\end{table*}

\subsection{BHT Detection using Sampled Marginal Posterior}

We perform the hypothesis test in \eqref{eq:eq2-7} by scaling it in
logarithm. Using the sampled marginal posterior  obtained from the
BP part, an nonparametric implementation of the hypothesis test in
\eqref{eq:eq2-7} is given as
\begin{align}\label{eq:eq3-21}
\log \frac{{\sum {{{\bf{r}}_1} \times {{\bf{p}}_{{X_i}|{\bf{Z}}}}}
}}{{\sum\limits {{{\bf{r}}_0} \times {{\bf{p}}_{{X_i}|{\bf{Z}}}}}
}}\mathop {\mathop \gtrless \limits_{{\mathcal{H}_0}}
}\limits^{{\mathcal{H}_1}} \log \frac{{1 - q}}{q}
\end{align}
where $\times$ is elementwise multiplication of vectors, and
$\mathbf{r}_0,\mathbf{r}_1 \in \mathbb{R}^{N_d}$ are reference
vectors from the signal prior knowledge,  defined as
\begin{eqnarray}\label{eq:eq3-22}
{{\bf{r}}_0}:=\frac{{{{\bf{p}}_{X_i|S_i =
0}}}}{{{{\bf{p}}_{X_i}}}},\,\,\,{{\bf{r}}_1}: =
\frac{{{{\bf{p}}_{X_i|S_i = 1}}}}{{{{\bf{p}}_{X_i}}}}.
\end{eqnarray}
This BHT-based detector is only compatible with the nBP approach
because the BHT detection requires full information on the
multimodally distributed posterior of $X_i$ which cannot be provided
through the pBP approach.

\subsection{Computational Complexity}
In our  uniform sampling-based nBP,  the density-messages are vectors with size
$N_d$. Therefore, the decoder requires $\mathcal{O}(LN_d)$ flops  to
calculate a VtF message $\mathbf{a}_{i \rightarrow j}^{(l)}$ and
$\mathcal{O}( {\frac{{NLN_d }}{M}\log N_d })$ flops for a FtV
message $\mathbf{b}_{j \rightarrow i}^{(l)}$ per iteration. In
addition, the cost of the FFT-based convolution given in
\eqref{eq:eq3-18} spends $\mathcal{O}(N_d\log N_d)$ flop if we
assume the row weight is $NL/M$ in average sense. Hence, the
per-iteration cost of the uniform sampling-based nBP is
$\mathcal{O}(NLN_d+M{\frac{{NLN_d }}{M}\log N_d })\approx
\mathcal{O}(NLN_d \log N_d)$ flops. For the BHT detection, the
decoder requires $\mathcal{O}(N_d)$ flops to generate the likelihood
ratio of \eqref{eq:eq3-21}, which is much smaller than that of the
BP part. Therefore, the cost for the BHT detection can be ignored.


For the linear MMSE estimation to find nonzeros on the support, the
cost can be reduced upto $\mathcal{O}(KM)$ flops by applying QR
decomposition \cite{Bjorck}.  Thus, the total complexity of the
proposed algorithm is $\mathcal{O}\left( l^* \times NLN_d \log N_d +
KM \right)$ flops and it is further simplified to $\mathcal{O}(l^*
\times N+KM )$ since $L$ and $N_d$ are fixed constants. In addition,
it is known that the message-passing process is applied recursively
until messages have been propagated along with every edge in the
tree-structured graph, and every signal element has received
messages from all of its neighborhood, which requires $l^* =
\mathcal{O}(\log N)$ iterations
\cite{CS-BP2},\cite{Mackey},\cite{Bishop}.
Therefore, we finally obtain $\mathcal{O}(N \log N
+KM)$ for the complexity of the proposed algorithm, BHT-BP.

\section{Performance Validation} \label{Numresult}
We validate performance of the proposed
algorithm, BHT-BP, with extensive experimental results.  Four  types of
experimental results are discussed in this section, as given below:
\begin{enumerate}
\item Threshold $(\frac{M}{N})^*$ for BP convergence,
\item Support detection performance over SNR,
\item MSE comparison to recent algorithms  over SNR,
\item Empirical calibration of BHT-BP over $N_d$ and $L$.
\end{enumerate}
The support detection performance is evaluated in terms of the
success rate of perfect support detection, defined as
\begin{align}\label{eq:eq2-24}
P_{\text{succ}}:=\Pr \{ \widehat{\mathbf{s}} =\mathbf{s} |
\mathbf{Z}=\mathbf{z} \},
\end{align}
and  the MSE comparison to the other algorithms is performed in
terms of normalized MSE, given as
\begin{eqnarray}\label{eq:eq2-245}
{\text{MSE}}: = \frac{ {\left\| {\widehat{\bf{x}}_0 - {\bf{x}}_0}
\right\|_2^2 } }{{\left\| {\bf{x}}_{0} \right\|_2^2 }}.
\end{eqnarray}
We generate all the experimental results by averaging the measures,
given  in \eqref{eq:eq2-24} and \eqref{eq:eq2-245}, with respect to
the signal $\mathbf{x}_0$ and the additive noise $\mathbf{w}$ using
Monte Carlo method\footnote{At every Monte Carlo trial, we realize
$\mathbf{x}_0$ and $\mathbf{w}$ to produce a measurement vector
$\mathbf{z}$ given the matrix $\mathbf{\Phi}$.}. In addition,  we
define a SNR measure used in the experiment as
\begin{eqnarray}\label{eq:eq2-25}
\text{SNR } :=10 \log_{10} \frac{{\mathbb{E}{\left\| {{\bf{\Phi}}
\mathbf{X}} \right\|_2^2 } }}{{M\sigma _{W }^2 }} \,\,\text{ (dB) }.
\end{eqnarray}

\begin{figure}[!t]
\centering
\includegraphics[width=8.5cm]{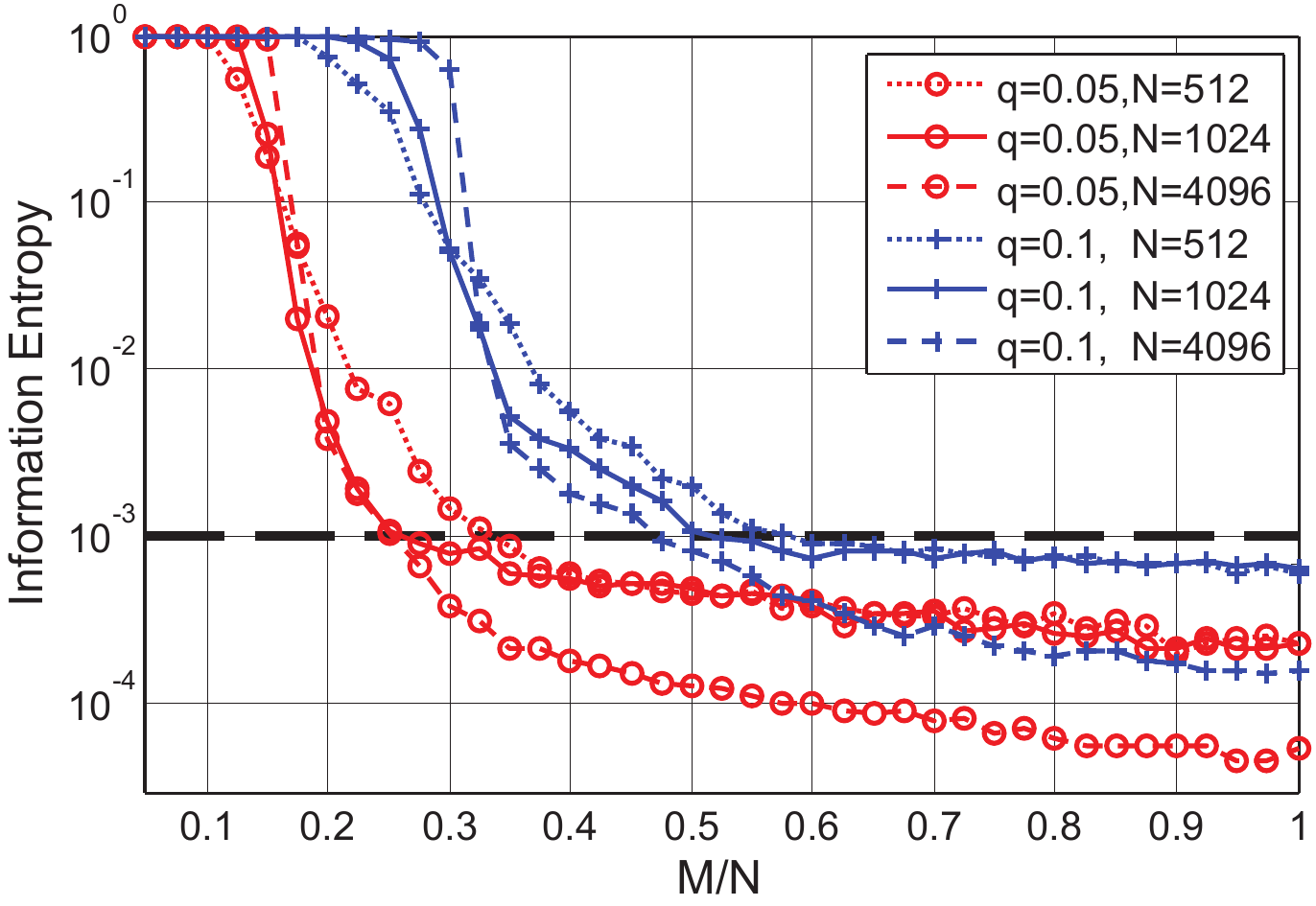}
\caption{The entropy phase transition curve over the sampling rate
$\frac{M}{N}$ for a variety of the signal length $N$ and the
sparsity rate $q$ where we set the threshold $(\frac{M}{N})^*$ to
the point achieving $\frac{1}{N}\sum\nolimits_{i = 1}^N
h(X_i|{\mathbf{Z}}= {\mathbf{z}}) \leq 10^{-3}$, which is given in
Table \ref{table2}. These curves are  information entropy of the
approximate marginal posterior,
$f_{\text{BP}({X_i})}^{(l)}(x|{\bf{Z}} = {\bf{z}})$, drawn with the
parameters  $\sigma_{X}=5$, $x_{min}=\sigma_{X}/4$, $N_d=256$,
$\varepsilon=10^{-5}, \lambda=10^{-4}$ and a noiseless setup. In
addition, we set the column weight of the matrix $\mathbf{\Phi}$ to
$L=4$ for $N=512, 1024$, and $L=5$ for $N=4096$, in this experiment.
} \label{PT}
\end{figure}

\begin{figure*}[!t]
\centering
\includegraphics[width=18cm]{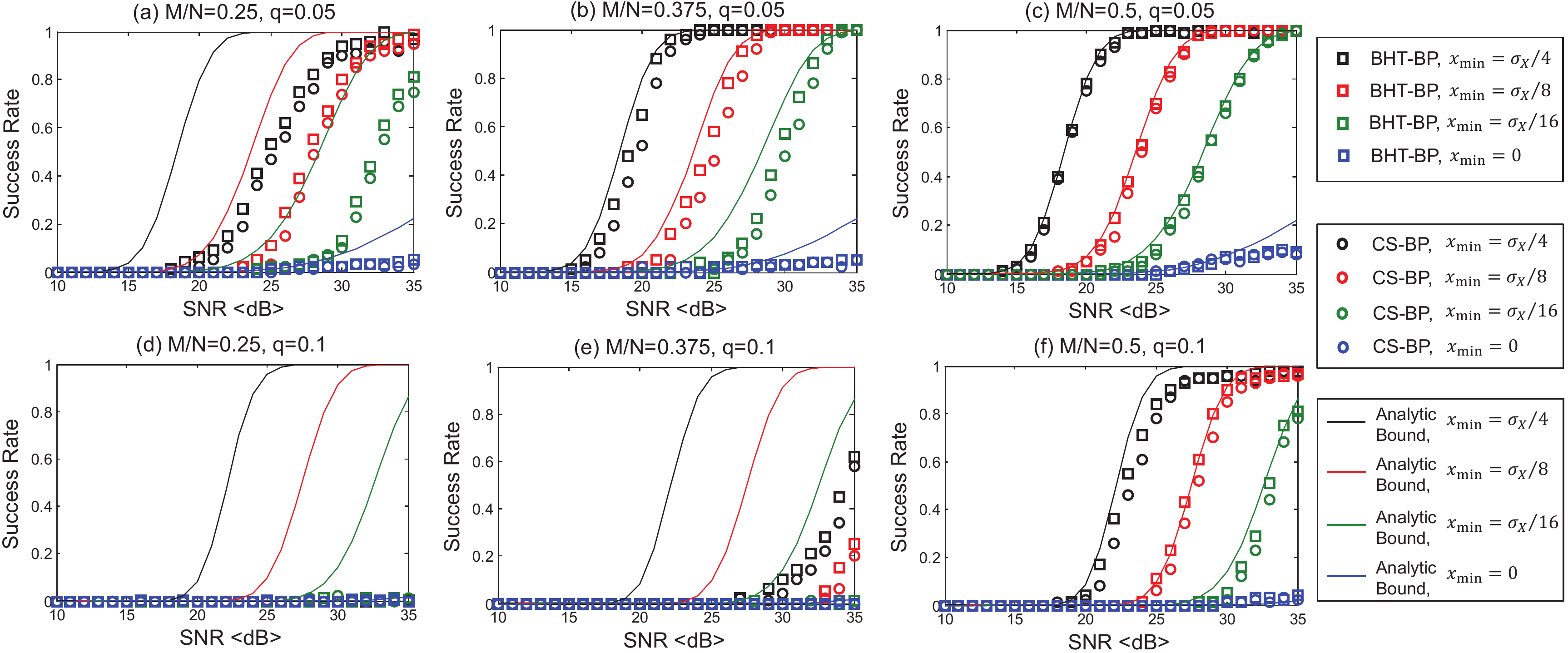}
\caption{Experimental result for the  success rate of support
detection over SNR for a variety of $x_{min}$ where we consider the
case of $N=1024$, $L=5$, and $\sigma_{X}=5$. In
Fig.\ref{fig:fig5-1}, we plot the success rate of of BHT-BP
(proposed) and CS-BP \cite{CS-BP2} together with the analytic bound
for the case of $\mathbf{\Phi} = \mathbf{I}$, where the nBP part of
the both algorithms is implemented with $N_d=256,
\varepsilon=10^{-5}, \lambda=10^{-4}$.}\label{fig:fig5-1}
\end{figure*}

For a comparison purpose, in this validation, we include several
recent Bayesian algorithms, CS-BP \cite{CS-BP2}, BCS \cite{BCS}, and
SuPrEM \cite{SuPrEM}, as well as  an $l_1$-norm based algorithm,
$l_1$-DS \cite{candes2}\footnote{The source codes of those
algorithms are obtained from each author's webpage. For CS-BP, we
implemented it by applying the uniform sampling-based nBP introduced
in Section IV-A.}. We provide brief introduction to the Bayesian
algorithms in Appendix I for interested readers. In this validation,
BHT-BP and CS-BP use the spike-and-dented slab prior, given in
\eqref{prior}, by applying the uniform sampling, \emph{i.e.},
$\mathbf{p}_{X_i} :=\text{Samp}\{f_{X_i}(x),T_s \}$.  Worth
mentioning here is  that the nBP-based solvers, such as  BHT-BP and
CS-BP, are only compatible with such an unusual signal prior, like
the spike-and-dented slab prior, which is one main advantage of the
nBP solvers. For the measurement matrix $\mathbf{\Phi}$, we
basically consider a LDPC-like matrix  in BHT-BP, CS-BP and BCS. In
case of SuPrEM, a LDF matrix is used for the measurement
generation\footnote{SuPrEM is only compatible with the LDF matrix
which was autonomously proposed in the work \cite{SuPrEM}.}, and
$l_1$-DS is performed with the standard Gaussian matrix as a
benchmark of the CS recovery. For fair compariosn, all types of the
matrices $\mathbf{\Phi}$ are equalized to have the same column
energy, \emph{i.e.} $\mathbb{E}\left[ {\left\| {{\phi
_{\text{column}}}} \right\|_2^2} \right] = L$; therefore, each entry
$\phi_{ji}$ of the standard Gaussian matrix is drawn from
$\mathcal{N}(\phi_{ji};0,\frac{L}{M})$. Table \ref{table1}
summarizes all the algorithms included in this performance
validation.

\subsection{Threshold $(\frac{M}{N})^*$ for BP Convergence,}
We claimed the  advantage of BHT-BP over CS-BP in support detection
with  the rate $\frac{M}{N}$ above a certain threshold
$(\frac{M}{N})^*$ in Section III-B. Given the rate $\frac{M}{N} \geq
(\frac{M}{N})^*$, a BP approximation of the marginal posterior
$f_{\text{BP}({X_i})}^{(l)}(x|{\bf{Z}} = {\bf{z}})$ contains
sufficiently less uncertainty on the true value $x_{0,i}$. We
empirically find the threshold $\left(\frac{M}{N}\right)^*$  in a
noiseless setup using the  average  information entropy,
$\frac{1}{N}\sum\nolimits_{i = 1}^N h(X_i|{\mathbf{Z}}=
{\mathbf{z}})$ which measures uncertainty of
$f_{\text{BP}({X_i})}^{(l)}(x|{\bf{Z}} = {\bf{z}})$.
The empirical entropy curves in Fig.\ref{PT} show sharp phase
transition as  $\frac{M}{N}$ increases. From the result, we set the
threshold to the point achieving $\frac{1}{N}\sum\nolimits_{i = 1}^N
h(X_i|{\mathbf{Z}}= {\mathbf{z}}) \leq 10^{-3}$, which is given in
Table \ref{table2}  for a variety of the signal length $N$ and the
sparsity rate $q$. We also note from Fig.\ref{PT} that the entropy
phase transition  becomes sharper as  $N$ increases.

\begin{table}
\footnotesize
\renewcommand{\arraystretch}{1.2}
\caption{Empirical  Threshold $(M/N)^*$ for the BP convergence}
\label{table2}
 \centering
\begin{tabular}{||c||c|c|c||}
\hline
Sparsity rate & $N=512$       & $N=1024$             & $N=4096$  \\
\hline \hline
$q=0.05$ &  0.325     &     0.25           &  0.25  \\
 \hline
$q=0.1$  &   0.575   &     0.50    &   0.475  \\
\hline
\end{tabular}
\end{table}

\subsection{ Support Detection Performance  over SNR}
Fig.\ref{fig:fig5-1} depicts an experimental comparison of the
success rate, defined in \eqref{eq:eq2-24}, between BHT-BP and CS-BP
over SNR for a variety of $x_{\min}$.
According to the threshold $(\frac{M}{N})^*$ given in Table \ref{table2}, the BP convergence is
achieved only for the cases of (a),(b),(c),(f) in Fig.\ref{fig:fig5-1}.
Therefore, we confine our discussion here to such cases,  claiming the advantage of
BHT-BP over CS-BP in support detection.

\subsubsection{SNR gain by BHT support detection }
The empirical results of Fig.\ref{fig:fig5-1} validate our claim
that BHT-BP has  more robust support detection ability against
noise, than CS-BP. Indeed, Fig.\ref{fig:fig5-1} shows that BHT-BP
enjoys a remarkable SNR gain from CS-BP in the low SNR regime. This
SNR gain is from difference of the detection criterion as discussed
in Section III-A. As SNR increases,  the success rate of the both
algorithms gradually approach to one. In the high SNR regime, BHT-BP
and CS-BP do not have notable difference in the performance.

\begin{figure}
\centering
\includegraphics[width=8.8cm]{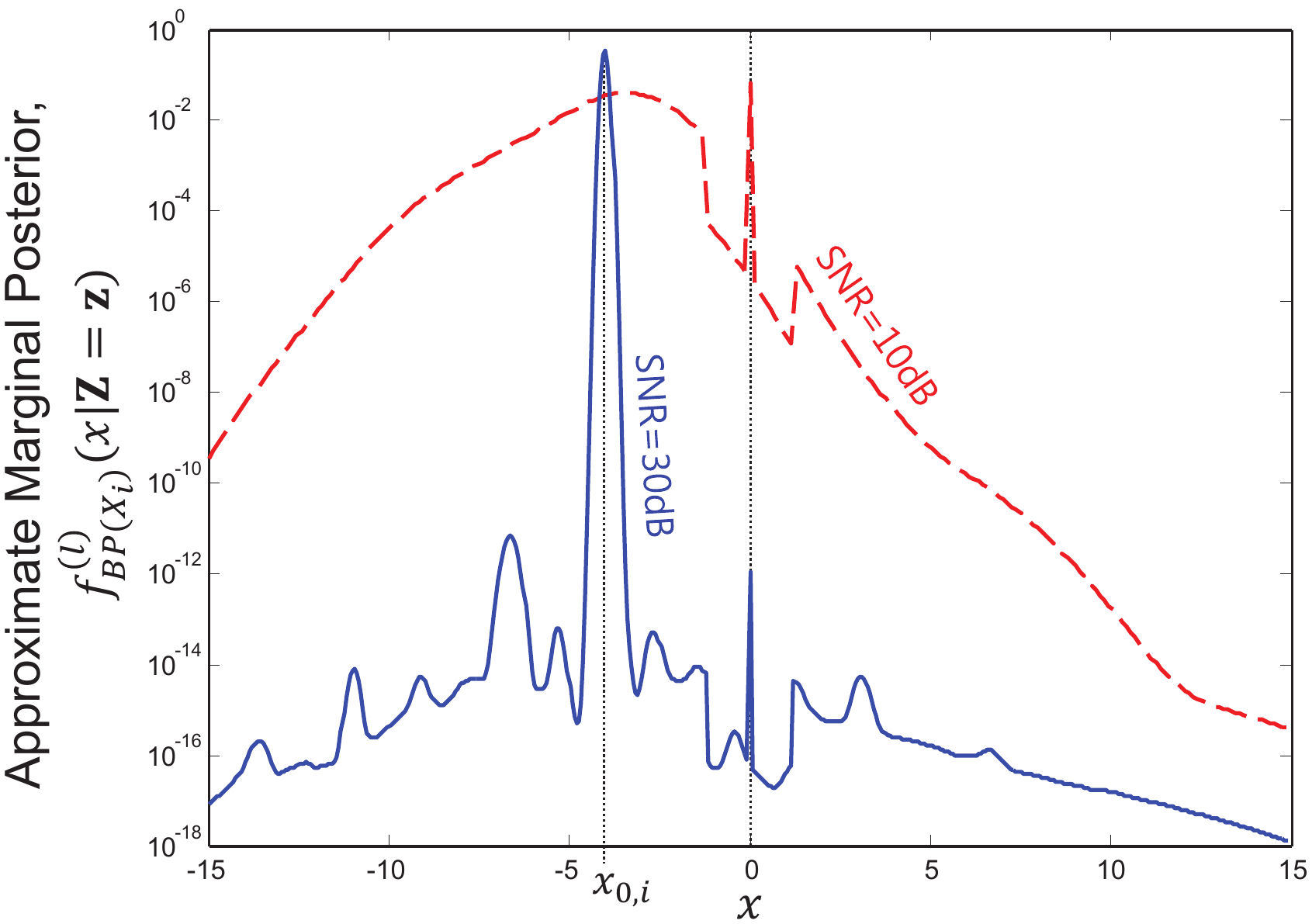}
\caption{Example of  the  approximate marginal  posterior
$f_{\text{BP}({X_i})}^{(l)}(x|{\bf{Z}} = {\bf{z}})$, obtained from
the nBP part, for two different SNRs: 10 dB and 30 dB, where the
true value of $X_i$ is $x_{0,i}=-4.0$, and the other parameters are
set to $M/N=0.5$, $q=0.05, \sigma_X=5$, $l^*=30$, and the minimum
value is $x_{min}=\sigma_X/4$.} \label{fig5-1supp}
\end{figure}

We support the advantage of BHT-BP over CS-BP with
Fig.\ref{fig5-1supp}. This figure depicts an exemplary marginal
posterior, obtained from the BP part, according to two different SNR
levels, SNR=10 and 30dB, where the true value of $X_i$ is
$x_{0,i}=-4.0$; hence $s_i=1$.
\begin{itemize}
\item When SNR is sufficiently high such as the SNR=30 dB case, both
of the algorithms can successfully detect the state $S_i$ from the
posterior since the probability mass is concentrated on the true
value $x_{0,i}$.
\item When SNR is low such as the SNR=10 dB case,
however, CS-BP may result in misdetection because the point-mass at
$x=0$ is higher than the point-mass at $x_{0,i}=-4.0$ due to the
additive noise, leading to $\widehat s_i=0$. In contrast, the BHT
detector decides the state $S_i$ by incorporating all the spread
mass due the noise. This is based on that the likelihoods
$f_{\mathbf{Z}}( \mathbf{z} |\mathcal{H}_{s_i})$, which construct
the hypothesis test of \eqref{BHT1}, is associated with the entire
range of the $x$-axis rather than a specific point-mass. Therefore,
BHT-BP can generate $\widehat s_i=1$ and success in the detection
even when SNR is low.
\end{itemize}

\subsubsection{Analytic Bound of BHT detection when $\mathbf{\Phi}=\mathbf{I}$}
Fig.\ref{fig:fig5-1} includes  an analytic bound of the BHT
detection for the case that the measurement matrix is an identity
matrix, \emph{i.e.} $\mathbf{\Phi}=\mathbf{I}$, such that there is
no performance degradation from  lack of measurements. Therefore,
this bound provides a performance benchmark  of the BHT detection
when $\frac{M}{N} \geq {\left( {\frac{M}{N}} \right)^*}$, because
exact marginal posteriors are given to the BHT detector under the
assumption of $\mathbf{\Phi}=\mathbf{I}$. We refer to Appendix II
for the detailed derivation of the analytic bound. This derivation
reveals that  the bound is  a function of  $q, x_{\min}$, and SNR.
In Fig.\ref{fig:fig5-1}, it is clearly shown that the empirical
points are fit into the analytic bounds as $\frac{M}{N}$ increases.

\subsubsection{Support detection with $x_{min}$}
Fig.\ref{fig:fig5-1} also shows the support detection behavior
according to $x_{min}$, confirming that $x_{min}$  is a key
parameter in the NSR problem. From Fig.\ref{fig:fig5-1}, we have the
observation as given in Note 3.\\
\\
\emph{ \bf{Note 3 (Empirical observations for $x_{\min}$):} }
\begin{itemize}
\item All the success rate curve shift toward high SNR region  as
$x_{min}$ decreases.
\item Extremely, when $x_{min}=0$, the experimental points
stay near zero even with $\frac{M}{N} \geq {\left( {\frac{M}{N}}
\right)^*}$ and high SNR.
\end{itemize}
These empirical observations intuitively tells us that contribution
of $x_{min}$ is as significant as SNR in the NSR problem,
implicating that we need SNR$\to \infty$ for the perfect support
recovery if the signal has $x_{min} \to 0$. Note that our
interpretation on the result here shows good agreement with not only
our analytic bound under the assumption of
$\mathbf{\Phi}=\mathbf{I}$, but also the information-theoretical
results \cite{Wainwright1},\cite{Wainwright2},\cite{Fletcher}
showing that support recovery is arbitrarily difficult by sending
$x_{min} \to 0$ even as SNR becomes arbitrarily large.

\subsection{MSE  Comparison to Recent Algorithms over SNR}
\begin{figure}[!t]
\centering
\includegraphics[width=8.8cm]{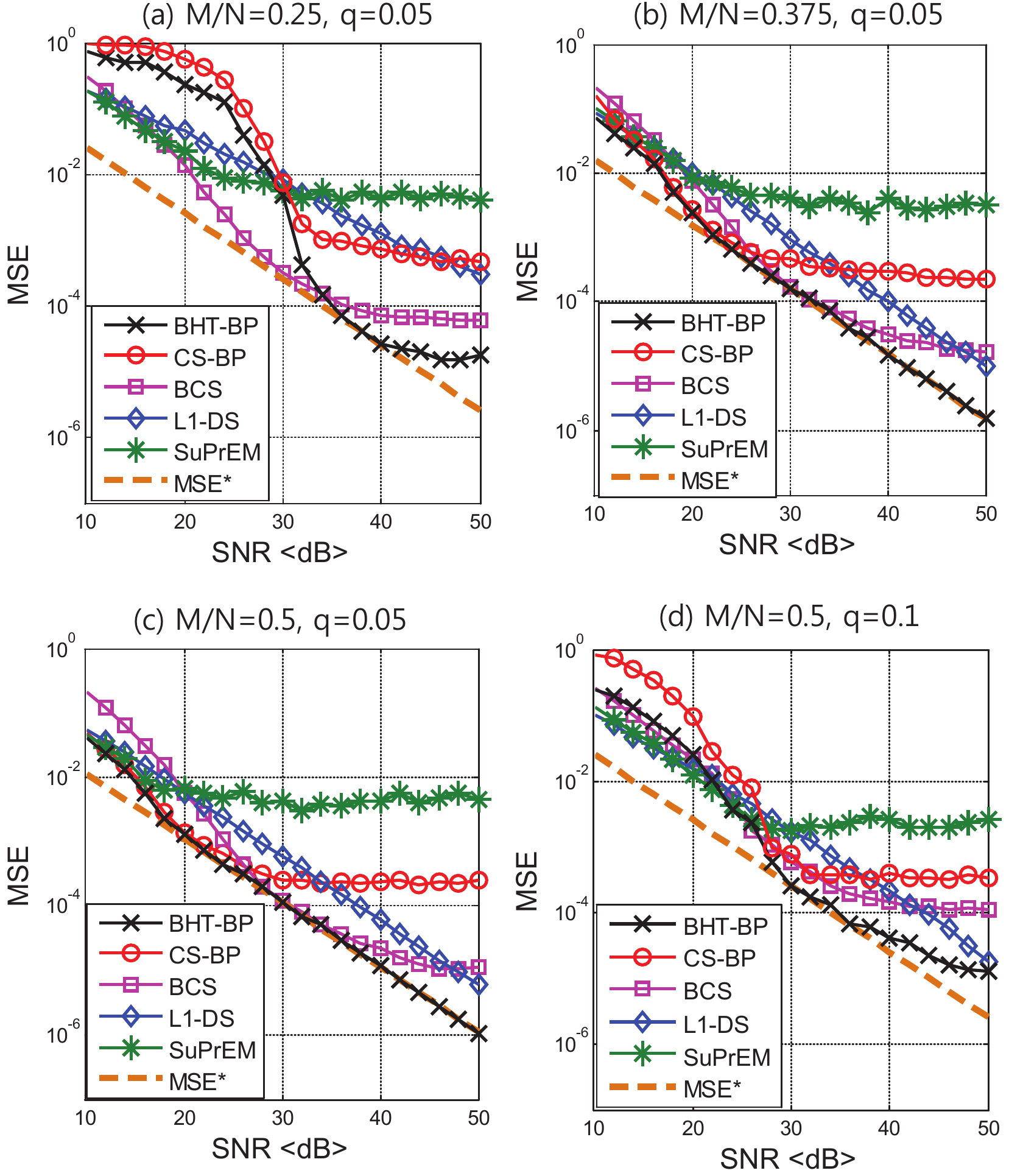}
\caption{MSE comparison among the algorithms (BHT-BP (Proposed),
CS-BP \cite{CS-BP2}, SuPrEM \cite{SuPrEM}, BCS \cite{BCS}, $l_1$-DS
\cite{candes2}) over SNR where we consider signal recovery with
$\frac{M}{N} \geq {\left( {\frac{M}{N}} \right)^*}$. We simulate the
MSE performance under $N=1024$, $L=5$, $\sigma_{X}=5$,
$x_{min}=\sigma_{X}/4$. The nBP part embedded in BHT-BP (proposed)
and CS-BP is implemented with $N_d=128$ and $\varepsilon=10^{-5},
\lambda=10^{-4}$.} \label{fig:fig5-2-1}
\end{figure}

\begin{figure}[!t]
\centering
\includegraphics[width=8.8cm]{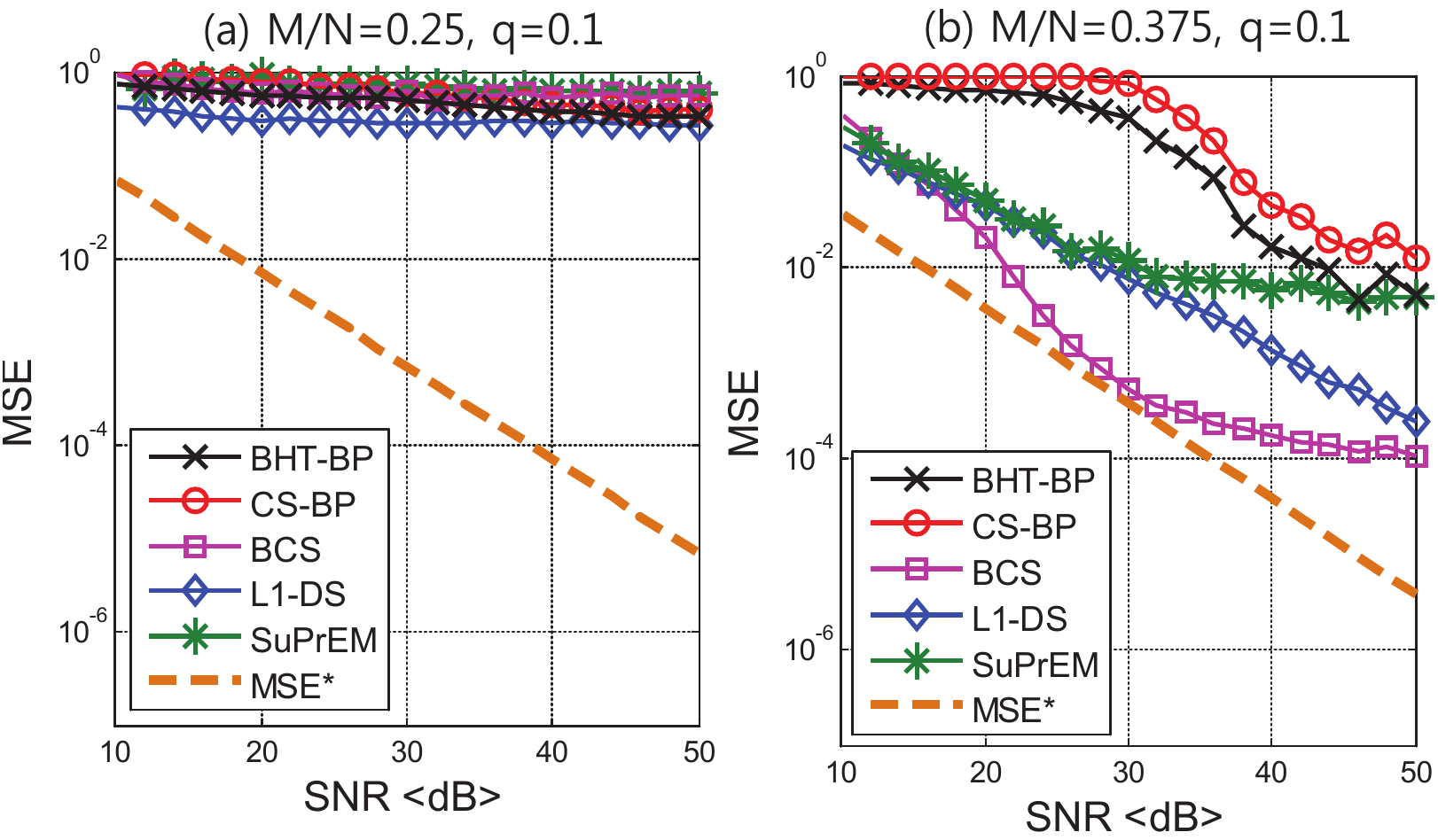}
\caption{MSE comparison among the algorithms (BHT-BP (Proposed),
CS-BP \cite{CS-BP2}, SuPrEM \cite{SuPrEM}, BCS \cite{BCS}, $l_1$-DS
\cite{candes2}) over SNR where we consider signal recovery with
$\frac{M}{N} < {\left( {\frac{M}{N}} \right)^*}$. We simulate the
MSE performance under $N=1024$, $L=5$, $\sigma_{X}=5$,
$x_{min}=\sigma_{X}/4$. The nBP part embedded in BHT-BP (proposed)
and CS-BP is implemented with $N_d=128$ and $\varepsilon=10^{-5},
\lambda=10^{-4}$.} \label{fig:fig5-2-2}
\end{figure}

In Fig.\ref{fig:fig5-2-1} and Fig.\ref{fig:fig5-2-2}, we provide an
MSE comparison among the algorithms listed in Table \ref{table1} and
the support-aware oracle estimator over SNR for a variety of
$(\frac{M}{N},q)$, where MSE$^*$ denotes the performance of the
support-aware oracle estimator, given as
\begin{eqnarray} \label{eq:eq5-28}
{\text{MSE}}^* := \frac{{{\rm{Tr}} \left[ { \left(\frac{1}{{\sigma
_{X}^2}}{\bf{I}} + \frac{1}{{\sigma _W^2}}{{\bf{\Phi
}}_{\mathbf{s}}^T {\bf{\Phi }}_{\mathbf{s}} } \right)^{ - 1} }
\right]}}{   {\mathbb{E} \left\| {{\bf{X}}} \right\|_2^2 }}.
\end{eqnarray}
 In this section, we discuss the comparison result by categorizing
the setup of $(\frac{M}{N},q)$ into two cases: the ``region of
$\frac{M}{N} \geq {\left( {\frac{M}{N}} \right)^*}$" and the
``region of $\frac{M}{N} < {\left( {\frac{M}{N}} \right)^*}$" cases,
according to the empirical threshold ${\left( {\frac{M}{N}}
\right)^*}$ given in Table \ref{table2}, where we fix the parameters
$N=1024$, $L=5$, $\sigma_{X}=5$, $x_{min}=\sigma_{X}/4$.

\subsubsection{MSE performance in region of $\frac{M}{N} \geq {\left( {\frac{M}{N}} \right)^*}$ }
With Fig.\ref{fig:fig5-2-1}, we argue that in the region of
$\frac{M}{N} \geq {\left( {\frac{M}{N}} \right)^*}$, BHT-BP catches
up with the oracle performance, MSE$^*$, beyond the SNR point
allowing the accurate support finding. Fig.\ref{fig:fig5-2-1}-(b)
and -(c) validate our claim by showing that the BHT-BP curve
coincides very closely with the MSE$^*$ curve beyond a certain SNR
point. Worth mentioning here is that the SNR point, which starts to
achieve the oracle MSE$^*$, nearly corresponds to the point which
attains the perfect support detection with $P_\text{succ} \approx
1.0$ in Fig.\ref{fig:fig5-1}. For the cases of
Fig.\ref{fig:fig5-2-1}-(a) and -(d), the BHT-BP curve does not fit
to the oracle MSE$^*$ at the high SNR region. The reason is coming
from lack of measurements for the BP convergence.  Indeed, it is
observed from Fig.\ref{PT} that the entropy points corresponding to
$(\frac{M}{N},q)$ of Fig.\ref{fig:fig5-2-1}-(a) and -(d) is in not a
steady region but a transient region. This means that the
corresponding posterior includes residual uncertainty  on $X_i$.
Although this residual uncertainty does not remarkably work in the
low SNR region due to noise effect, it is gradually exposed as SNR
increases, degrading the MSE performance in the high SNR region.

In Fig.\ref{fig:fig5-2-1}, the CS-BP curve  forms an
error floor as SNR increases, leading to a MSE gap from BHT-BP in the high SNR regime.
This MSE gap is mainly caused by the quantization error of the nBP. Since CS-BP obtains its estimate directly from
the sampled posteriors, the quantization error is
unavoidable, leading to an error floor. The level of the floor
can be approximately predicted by the MSE degradation of the quantization, given as
\begin{align}\label{eq:eq5-28}
\frac{{\mathbb{E}\left\| {{Q_{{T_s}}}[{{\bf{X}}_{\bf{S}}}] -
{{\bf{X}}_{\bf{S}}}}
\right\|_2^2}}{{\mathbb{E}||{{\bf{X}}_{\bf{S}}}||_2^2}} =
\frac{{T_s^2/12}}{{\sigma _X^2}}=\frac{3}{N_d^2}
\end{align}
where $Q_{T_s}[\cdot]$ is the quantization function with the step
size $T_s$ given in \eqref{eq:eq3-14}, and $\mathbf{X}_{\mathbf{S}}$
is a random vector on the signal support ${\mathbf{S}}$.
Under our joint detection-and-estimation
structure, the LMMSE estimator \eqref{eq:eq2-10} enables BHT-BP to
go beyond the error floor.

For SuPrEM, the performance is poor to the other algorithms in
the experimental results of Fig.\ref{fig:fig5-2-1}. But, it is not
surprising since SuPrEM is basically for signals having fixed signal
sparsity $K$.\footnote{We empirically confirmed that when $K$ is
fixed, SuPrEM works as comparable to BHT-BP even though we does not
include that result in this paper.} Indeed, the SuPrEM algorithm requires the sparsity $K$
as an input parameter. However, in many cases, the
signal sparsity $K$ is unknown and random. In our basic setup,
recall that we assumed signals having Binomial random sparsity,
\emph{i.e.}, $K \sim \mathcal{B}(k;N,q)$. Therefore, naturally
SuPrEM underperforms the other algorithms in this experiment.
$l_1$-DS and BCS are comparable to BHT-BP, but $l_1$-DS has a
certain SNR loss from the BHT-BP over all range of SNR, and BCS
shows an error floor at high SNR region.

\subsubsection{MSE performance in region of $\frac{M}{N} < {\left( {\frac{M}{N}} \right)^*}$ }
In Fig.\ref{fig:fig5-2-2}, we investigate the MSE comparison in  the
region of $\frac{M}{N} < {\left( {\frac{M}{N}} \right)^*}$.
 Under the setup of
$\frac{M}{N}=0.25, q=0.1$, every algorithm generally does not work as shown in Fig.\ref{fig:fig5-2-2}-(a).
From the setup of $\frac{M}{N}=0.375, q=0.1$, all the algorithms begin to find signals but,
 BHT-BP  underperforms BCS, L1-DS, and SuPrEM in this setup,
as shown in Fig.\ref{fig:fig5-2-2}-(b). The reason is that in the region of $\frac{M}{N} < {\left( {\frac{M}{N}} \right)^*}$,
the BP does not converge properly due lack of the measurements such that  probability mass on the true value $x_{0,i}$ is not dominant in the
approximate marginal posteriors, as we discussed in Section III-B.
From the results, we conclude that BHT-BP is not advantageous over the other algorithms excluding CS-BP
 when sufficient measurements is not maintained for the signal sparsity.

\begin{figure}
\centering
\includegraphics[width=8.8cm]{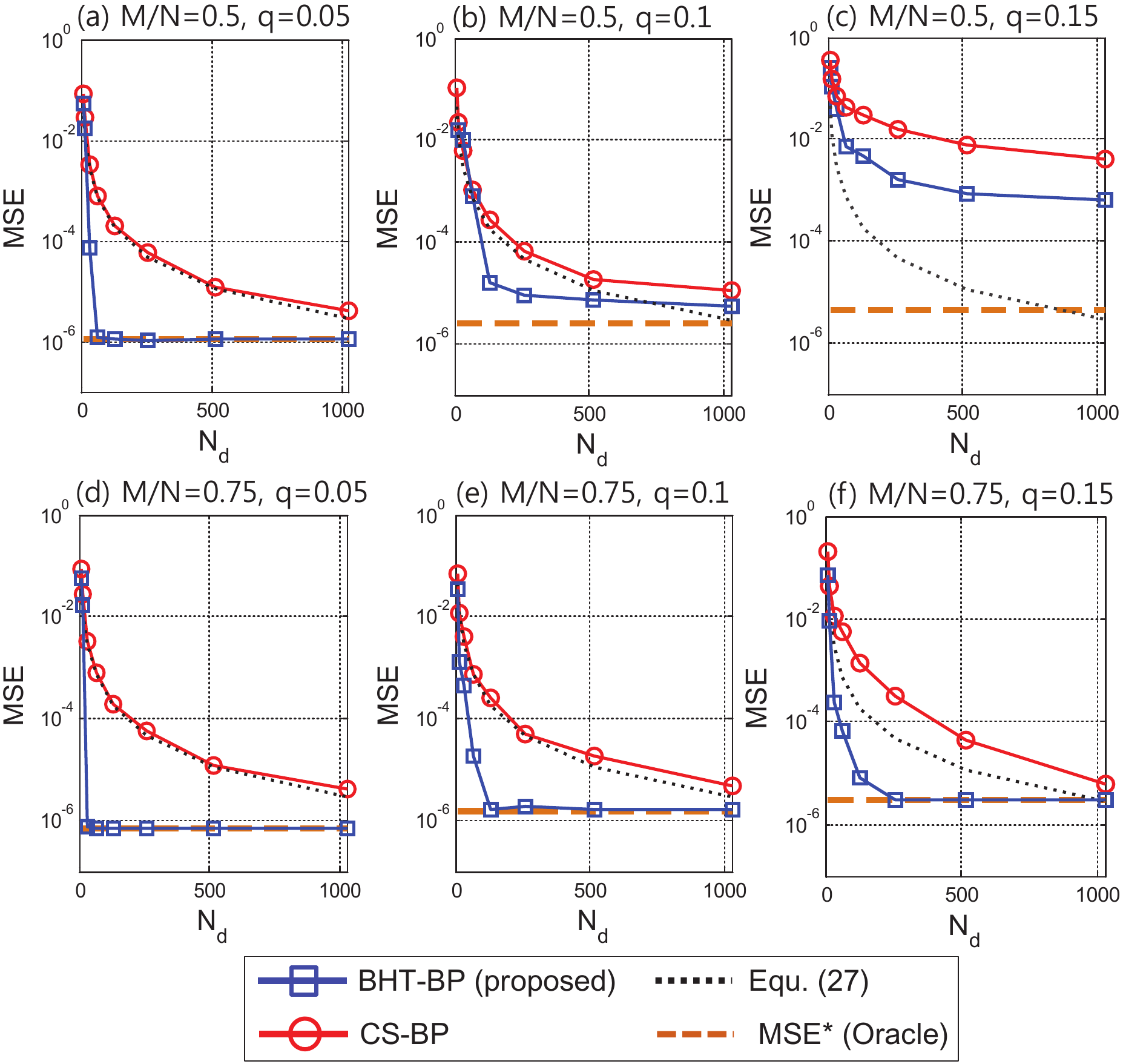}
 \caption{MSE comparison of BHT-BP (proposed) and CS-BP \cite{CS-BP2}
 with clean measurements  (SNR=50dB) over $N_d\in \{8, 16, 32, 64, 128, 256, 512, 1024\}$ for a variety of $\frac{M}{N},q$, where we plot the MSE curves together with the MSE degradation by the
quantization error, given in \eqref{eq:eq5-28}. In this experiment,
we consider a case of $N=1024, L=5, x_{min}=\sigma_{X}/4,
\varepsilon=10^{-5}, \lambda=10^{-4}$. } \label{fig:Fig3-2}
\end{figure}

\subsection{Empirical Calibration of BHT-BP over $N_d$ and $L$}
\subsubsection{Number of samples $N_d$ for nBP }
From the discussion in Section IV-C, one can argue that the
complexity of BHT-BP is highly sensitive to the number of samples
$N_d$; therefore, BHT-BP cannot be low-computational in a certain
case. It is true, but we claim that the effect of $N_d$ is limited
in the BHT-BP recovery. To support our claim, Fig.\ref{fig:Fig3-2}
compares MSE performance of BHT-BP, CS-BP, and the support-aware
oracle estimator as a function of  $N_d$ in a clean setup (SNR=50
dB) where we plot the MSE curves together with the MSE degradation
by the quantization error, given in \eqref{eq:eq5-28}. From
Fig.\ref{fig:Fig3-2}, we confirm that BHT-BP can achieve the oracle
performance if $N_d$ is beyond a certain level and $(\frac{M}{N},q)$
belongs to the success phase, whereas CS-BP cannot provide the
oracle performance even as $N_d$ increases. Consequently, $N_d$ does
not significantly contribute to the MSE of the BHT-BP recovery once
$N_d$ exceeds a certain level. This result implies that the
complexity of the BHT-BP recovery can be steady with a constant
$N_d$ in practice. Therefore, BHT-BP can holds the low-computational
property given by the BP philosophy. In addition, we confirm from
Fig.\ref{fig:Fig3-2} that the MSE of CS-BP is bounded by
\eqref{eq:eq5-28}.

\subsubsection{Column weight  $L$ of LDPC-like matrices }
\begin{figure}
\centering
\includegraphics[width=8.8cm]{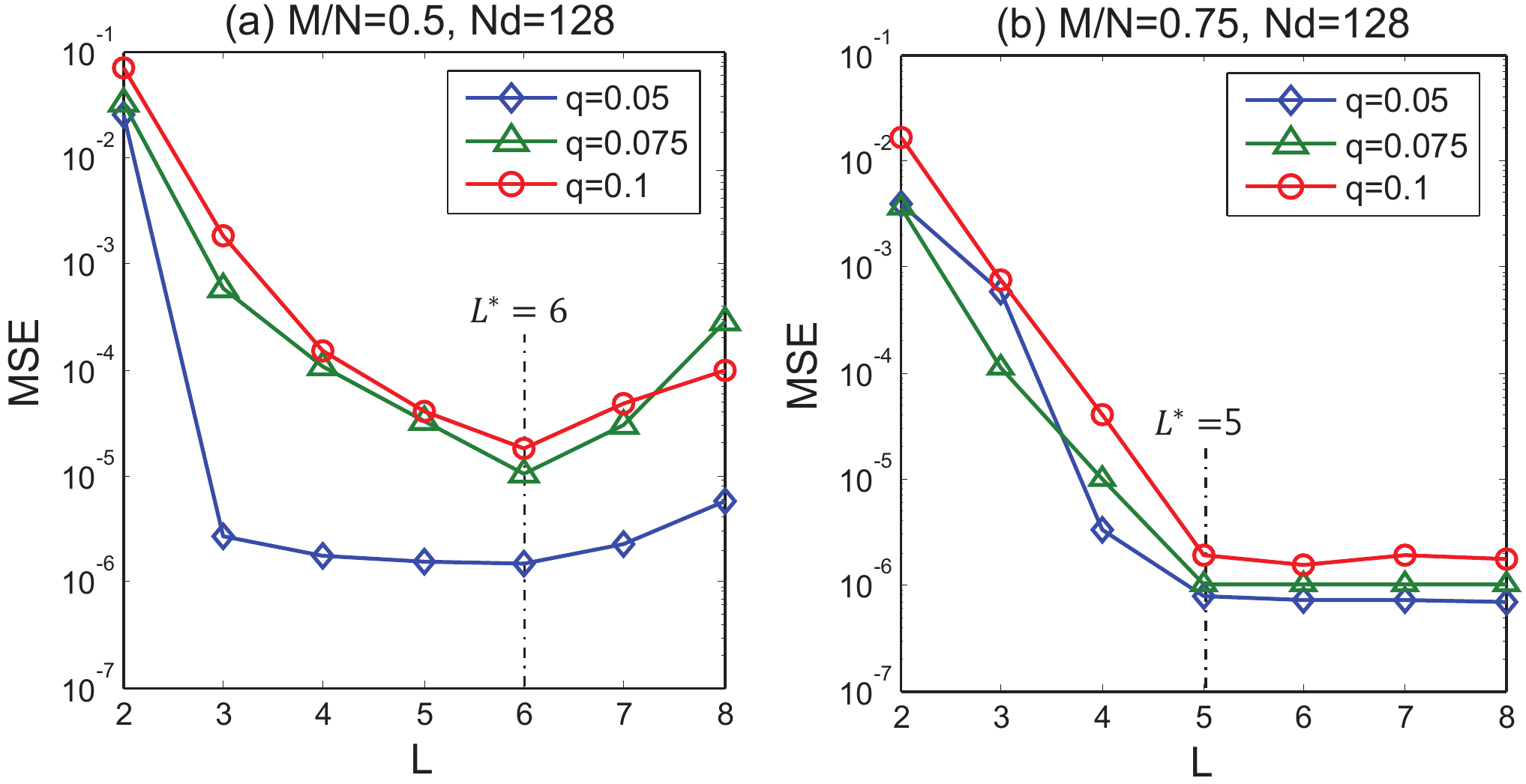}
\vspace{-5pt} \caption{MSE performance of the BHT-BP recovery over
the column weight $L$ of the measurement matrix $\mathbf{\Phi}$
where we set $N=1024, N_d=128, x_{min}=\sigma_{X}/4$,
$\varepsilon=10^{-5}, \lambda=10^{-4}$ and consider the recovery
from clean measurements (SNR= 50dB). } \label{fig:fig3-3}
\end{figure}

Another interesting question is  how to determine the column weight
$L$ of the LDPC-like matrix $\mathbf{\Phi}$ for BHT-BP.
Fig.\ref{fig:fig3-3} provides an answer for this question by showing
the MSE of the BHT-BP recovery as a function of $L$, where we
consider the recovery from clean measurements (SNR= 50 dB). When
$\frac{M}{N}$ is sufficiently large, for example $\frac{M}{N}=0.75$ as shown in
Fig.\ref{fig:fig3-3}-(b), the BHT-BP recovery generally becomes
accurate as $L$ increases. Then, the accuracy is almost constant
after a certain point $L=L^*$. On the other hands, when $\frac{M}{N}$ is not
sufficient, for example $\frac{M}{N}=0.5$ as shown in
Fig.\ref{fig:fig3-3}-(a), the recovery accuracy rather can be
degraded beyond a certain point $L^*$. The reason is that when $\frac{M}{N}$
is small, the large $L$ spoils the tree-structured property of the matrix
$\mathbf{\Phi}$, reducing the accuracy of the marginal
posterior approximation  by BP \cite{Mackey},\cite{Richardson}. Therefore, $L$ should
keep as small as possible once the desirable recovery accuracy is
achieved. In the case of Fig.\ref{fig:fig3-3},  we empirically set
$L^*= 6,5$ for $\frac{M}{N}=0.5,0.75$ respectively.

From the calibration shown in Fig.\ref{fig:Fig3-2} and
Fig.\ref{fig:fig3-3}, we support our claim that the computational
cost of BHT-BP can be $\mathcal{O}(N \log N + KM)$ in practice by
fixing $L$ and $N_d$, as we discussed in Section IV-C.

\section{Conclusion}
The theoretical and empirical research in this paper demonstrated
that BHT-BP is powerful as not only  a low-computational solver, but
also  a noise-robust solver. In BHT-BP, we employed a joint
detection-and-estimation structure consisting of the BHT support
detection and the LMMSE estimation for the nonzeros on the signal
support. We have shown that the BHT-BP detects the signal support
based on a sequence of binary hypothesis tests, which is related to
the criterion of the minimum detection error probability. This
support detection approach brings SNR gain of BHT-BP from CS-BP
\cite{CS-BP2}, which is an existing nBP-based algorithm, for the
support detection. In addition, we noted the fact that BHT-BP
effectively removes the quantization error of the nBP approach in
the signal recovery. We have claimed that our joint
detection-and-estimation strategy prevents from degrading the MSE by
the  quantization error. We have supported the claim based on an
empirical result that the performance of BHT-BP achieves the oracle
performance when sufficient measurements is maintained for the
signal sparsity. Furthermore, we confirm the impact of $x_{min}$ on
the noisy sparse recovery (NSR) problem via BHT-BP. Based on the
empirical evidence, we showed that exact sparse recovery with small
$x_{min}$ is very demanding  unless sufficiently large SNR is
provided, which is an agreement with the result of
\cite{Wainwright1},\cite{Wainwright2},\cite{Fletcher} that
emphasizes the importance of $x_{min}$ in the NSR problem.

\section*{Appendix I\\
Brief Introduction to \\Recent Bayesian Algorithms}  In this
appendix, we provide a brief introduction to some previously
proposed Bayesian algorithms for the NSR problem: BCS \cite{BCS},
CS-BP \cite{CS-BP2}, SuPrEM \cite{SuPrEM}. These algorithms have
been developed by applying several types of signal prior PDFs and
statistical techniques. These algorithms are included for
simulation based comparison in Section V.

\subsection{BCS Algorithm}
Ji \emph{et al.} proposed a Bayesian algorithm based on the
\emph{sparse Bayesian learning} (SBL) framework, called BCS
\cite{BCS}. In the SBL framework,  a  two-layer hierarchical
Gaussian model has been invoked for signal estimation. Namely, the
signal prior PDF takes the form of
\begin{eqnarray}
{f_{\mathbf{X}}}({\bf{x}}|a,b) = \prod\limits_{i = 1}^N
{\int_0^\infty {\mathcal{N}({x}_i;0,\gamma _i^{ - 1})}  f_{\Gamma}
({\gamma_i}|a,b)d{\gamma_i }},
\end{eqnarray}
where $f_{\Gamma} ({\gamma_i }|a,b)$ is a hyper-prior following the
Gamma distribution with its parameters $a,b$. Then, the MAP estimate
$\widehat{\mathbf{x}}_0$ of the signal can be analytically expressed
as a function of the hyperparameter
$\mathbf{\Gamma}=[\gamma_1,...,\gamma_N]$, the measurement matrix
$\mathbf{\Phi}$, and the noisy measurements $\mathbf{z}$.

In BCS, the hyperparameter $\mathbf{\Gamma}$ is estimated by
performing a type-II \emph{maximum likelihood} (ML) procedure
\cite{SBL}. Specifically, the type-II ML finds the hyperparameter
$\mathbf{\Gamma}$ maximizing the evidence PDF, \emph{i.e.},
${f_{\mathbf{Y}}}({\mathbf{y}}|{\mathbf{\Gamma }}) = \int\limits
{{f_{\mathbf{Y}}}({\mathbf{y}}|{\mathbf{X}} =
{\mathbf{x}}){f_{\mathbf{X}}}({\mathbf{x}}|} {\mathbf{\Gamma
}})\,d{\mathbf{x}}$. The expectation maximization (EM) algorithm can
be an efficient approach for the type-II ML procedure. The strategy
of EM is to derive a lower bound on the log evidence PDF, $\log
{f_{\mathbf{Y}}}({\mathbf{y}}|{\mathbf{\Gamma }})$, at the E-step,
and  optimize that lower bound to find $\mathbf{\Gamma}$ at the
M-step. The E-step and M-step are iterated until the  lower bound
becomes tighter.

The BCS algorithm is input parameter-free, which means this
algorithm is adaptive to any types of signals and noise level since
BCS properly catches the hyperparameter $\mathbf{\gamma}$ and the
noise variance $\sigma_W^2$  during the recovery. In addition, the
BCS algorithm is well compatible with any type of the measurement
matrices.

\subsection{CS-BP Algorithm}
Baron \emph{et al.} for the first time proposed the use of BP to the
sparse recovery problem with LDPC-like measurement matrices
\cite{CS-BP2}. The algorithm is called CS-BP. Signal model of CS-BP
is a compressible signal which has a small number of large elements
and a large number of near-zero elements. The authors associated
this signal model with a two-state mixture Gaussian prior, given as
\begin{eqnarray}\label{eq:eq_4}
{f_{\mathbf{X}}}(\mathbf{x}) = \prod\limits_{i = 1}^N { \left[
q\mathcal{N}(x_i;0,\sigma _{{X_1}}^2) + (1 -
q)\mathcal{N}(x_i;0,\sigma _{{X_0}}^2) \right] },
\end{eqnarray}
where $q \in [0,1)$ denotes the probability that an element has the
large value, and $\sigma_{X_1}  \gg \sigma_{X_0}$. Therefore, the
prior is fully parameterized with $\sigma_{X_0},\sigma_{X_1}$, and
$q$. CS-BP performs MAP or MMSE estimation using marginal posteriors
obtained from BP similarly to the proposed algorithm, where the
authors applied both nBP and pBP approaches for the BP
implementation. The recovery performance is not very good when
measurement noise is severe since the CS-BP was basically designed
to work under noiseless setup.

\subsection{SuPrEM Algorithm}
Most recently, Akcakaya \emph{et al.} proposed SuPrEM under a
framework similar to BCS which uses the two-layer hierarchical
Gaussian model for the signal prior. SuPrEM was developed under the
use of a specific type of hyper-prior called the Jeffreys' prior
$f_{\mathcal{J}}(\tau_i) =1/ \beta_i, \beta_i \in [T_{i},\infty]
\,\,\forall i \in \mathcal{V}$. This hyper-prior reduces the number
of input parameters while sparsifying the signal. The overall signal
prior PDF is given as
\begin{eqnarray}\label{eq:eq_5}
{f_{\mathbf{X}}}({\bf{x}}) = \prod\limits_{i = 1}^N {\int_0^\infty
{\mathcal{N}({x}_i;0,\beta _i)}  f_{\mathcal{J}}
({\beta_i})d{\beta_i }}.
\end{eqnarray}
SuPrEM utilizes the EM algorithm to find each hyperparameter
$\beta_i$ like the BCS algorithm. However,  differently from BCS
that calculates the signal estimate $\widehat{\mathbf{x}}_0$ using
matrix operations which include matrix inversion, SuPrEM
elementwisely calculates the signal estimate from $\beta_i$ via a
pBP algorithm. Therefore, SuPrEM can be more computationally
efficient than BCS.

The measurement matrix used in SuPrEM is restricted to an LDPC-like
matrix which has fixed column and row weights, called
low-density-frames (LDF). They are reminiscent of the regular LDPC
codes \cite{Gallager}. In addition, the signal model is confined to
$K$-sparse signals consisting of $K$ nonzeros and $N-K$ zeros since
SuPrEM includes a sparsifying step which chooses the $K$ largest
elements at each end of iteration. The noise variance $\sigma_W^2$
is an optional input to the algorithm. Naturally, if the noise
variance is provided, SuPrEM will produce an improved recovery
performance.

\section*{Appendix II\\
Success Rate Analysis of the BHT  Detection \\when
$\mathbf{\Phi}=\mathbf{I}$}

Under the assumption of $\mathbf{\Phi}=\mathbf{I}$, the measurement
channel can be decoupled to $N$ scalar Gaussian channels which are
$Z_j=X_{i} + W_j\,\,\,\forall i,j\in \mathcal{V}$ where clearly
$i=j$ holds. Accordingly, the success rate, given in
\eqref{eq:eq2-24}, can be represented as the product of the
complementary probability of the state error rate (SER) given in
\eqref{SER}, \emph{i.e.}, ${P_\text{succ}} = {(1 -
{P_\text{SER}})^N}$. Then, the problem is reduced to the analysis of
the rate ${P_{\text{SER}}}$ (see Fig.\ref{fig3-5}). The conditional
SER given the hypothesis $\mathcal{H}_{s_i}$ is calculated from the
likelihood PDF ${f_{{Z_j}}}(z|{\mathcal{H}_{{s_i}}})$ as following:
\begin{align}\label{condSER}
{P_{{\text{SER}}|{{\mathcal{H}}_{{s_i}}}}}: =& \Pr \{ {\widehat s_i}
\ne {s_i}|{\mathcal{H}_{{s_i}}}\}  = \int_{{ \overline
D_{{\mathcal{H}_{{s_i}}}}}}
{{f_{{Z_j}}}(z|{\mathcal{H}_{{s_i}}})dz},
\end{align}
where we define the decision regions with a threshold $\gamma'$ as
\begin{align}\label{decisionregion}
{D_{{\mathcal{H}_0}}}: = \{ |z| <   \gamma'\}\,\, \text{and}\,\,
D_{{\mathcal{H}_1}}: = \{ |z| \geq  \gamma'  \},
\end{align}
and $\overline D _{{\mathcal{H}_0}} = {D_{{\mathcal{H}_1}}}$ vice
versa. The likelihood PDFs can be obtained from
\begin{align}
{f_{{Z_j}}}(z|{\mathcal{H}_{{s_i}}})={\int {
{{f_{Z_j}}(z|{X_i} = x)} {f_{{X_i}}}( x|{\mathcal{H}_{s_i}}) dx} }
\end{align}
as we have done in \eqref{BHT1},  where ${{f_{Z_j}}(z|{X_i} =
x)}=\mathcal{N}(z;x,\sigma_W^2)$ under the scalar Gaussian channel.
Then, the likelihood given ${\mathcal{H}_0}$ simply becomes
${f_{{Z_j}}}(z|{\mathcal{H}_0}) = \mathcal{N}(z;0,\sigma _W^2)$. In
contrast, the likelihood conditioning ${\mathcal{H}_1}$ is not
straightforward due to the dented slab part of our prior in
\eqref{prior}, which is given by
\begin{align}\label{likelihoodPDF2}
 & {f_{{Z_j}}}(z|{ \mathcal{H}_1})  \propto   \int_{|x| \geq x_{\min} } {\mathcal{N}(z;x,\sigma _W^2)\mathcal{N}(x;0,\sigma _X^2)dx} \nonumber\\
 &\,\,\,\,\,\,\,\,\,\,\,\,\,\,\,\,\,\,\,\,\,\,\,\,\,\,\,\,\,\,\,\,\,\,+\lambda\int_{ |x|<x_{\min}} {\mathcal{N}(z;x,\sigma _W^2)dx}\nonumber\\
   &\,\,\,\,\,\,\,\,\,\,\,\,\,\,\,\,\,\,\,\,\,\,\,\,\,\,\,= \mathcal{N}(z;0,\sigma _W^2 + \sigma _X^2) \\
&\,\,\,\,\,\,\,\,\,\,\,\,\,\,\,\,\,\,\,\,\,\,\,\,\,\,\,\,\,\,\,\,\,\,\,\times
\begin{array}{l}
   \left(1 - \frac{1}{2} \text{erf}\left(\frac{{A(z)}}{{\sqrt 2 }}\right) - \frac{1}{2}\text{erf}\left(\frac{{B(z)}}{{\sqrt 2
   }}\right)\right)
   \end{array}\nonumber\\
  &\,\,\,\,\,\,\,\,\,\,\,\,\,\,\,\,\,\,\,\,\,\,\,\,\,\,\,\,\,\,\,\,\,\,\,+ \frac{\lambda }{2}
  \begin{array}{l}\left( {{\text{erf}}\left( {\frac{{{x_{{\text{min}}}} - z}}{{{\sigma _W}\sqrt 2 }}} \right) + \text{erf}\left( {\frac{{{x_{{\text{min}}}} + z}}{{{\sigma _W}\sqrt 2 }}} \right)} \right)\end{array}\nonumber
\end{align}
where normalization is required to satisfy $\int {f_{{Z_j}}}(z|{ \mathcal{H}_1}) dz= 1$, and the functions
$A(z), B(z)$ are respectively described as
\begin{align*}
&A(z): = \frac{{{x_{\min }}\left(\frac{1}{{\sigma _W^2}} +
\frac{1}{{\sigma _X^2}}\right) - \frac{z}{{\sigma _W^2}}}}{{\sqrt
{\frac{1}{{\sigma _W^2}} + \frac{1}{{\sigma _X^2}}} }},\,\, \\
&B(z):
= \frac{{{x_{\min }}\left(\frac{1}{{\sigma _W^2}} + \frac{1}{{\sigma
_X^2}}\right) + \frac{z}{{\sigma _W^2}}}}{{\sqrt {\frac{1}{{\sigma
_W^2}} + \frac{1}{{\sigma _X^2}}} }}.
\end{align*}
In this problem, an analytical expression of $\gamma'$ is
unattainable from the equality condition \eqref{eqcondition} since
the PDF ${f_{Z_j}}(z|{\mathcal{H}_1})$ involves the error function
terms as shown in \eqref{likelihoodPDF2}. Therefore, we utilize a
root-finding algorithm to compute $\gamma'$.  We use the SNR
definition given in \eqref{eq:eq2-25} such that ${\text{SNR =
}}10{\log _{10}}\frac{{qL\sigma _X^2}}{{\sigma _W^2}}$ under the
assumption of $\mathbf{\Phi}=\mathbf{I}$. We specify the decision
regions \eqref{decisionregion} with $\gamma'$,  finalizing this
analysis by computing the condition SERs, which are given as
\begin{align}
 {P_{{\text{SER}}|{{\mathcal{H}}_0}}} &= {1 -
{\text{erf}}\left( {\frac{{{ \gamma' }}}{{ {\sigma _W}\sqrt {2  }}}}
\right)},\label{condSER0} \\
 {P_{{\text{SER}}|{{\mathcal{H}}_1}}} &=2\int_0^{\gamma'} {f_{{Z_j}}}(z|{ \mathcal{H}_1}) {dz}\label{condSER1}
\end{align}
where the calculation of ${P_{{\text{SER}}|{{\mathcal{H}}_1}}}$
requires a numerical integration owing to the error function terms
in ${f_{{Z_j}}}(z|{ \mathcal{H}_1})$. Using \eqref{SER},
\eqref{condSER0}, and \eqref{condSER1}, we can evaluate the SER,
then obtaining the success rate of the BHT detection when
$\mathbf{\Phi}=\mathbf{I}$.  We  compare this analysis result to the
empirical results  in Section V-B.

%
\newpage

\begin{IEEEbiography}[{\includegraphics[width=1in,height=1.25in,clip,keepaspectratio]{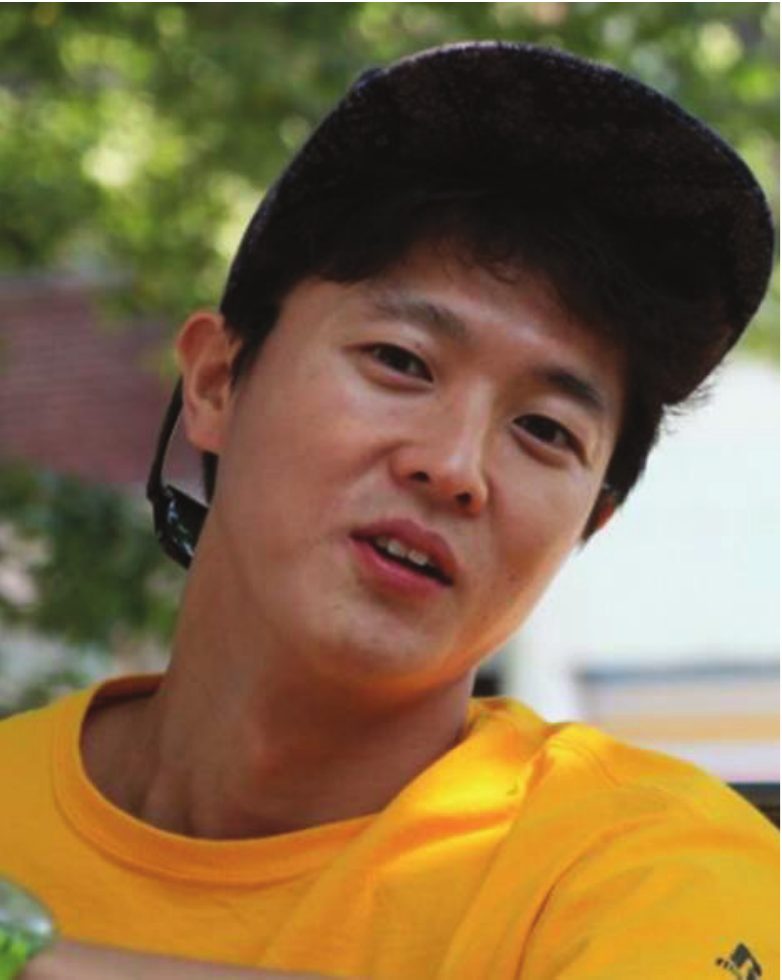}}]{Jaewook Kang}
(M'14) received the B.S. degree in information and communications
engineering  (2009) from Konkuk University, Seoul, Republic of
Korea, and the M.S. degree in information and communications
engineering (2010) from the Gwangju Institute of Science and
Technology (GIST), Gwangju, Republic of Korea. He is currently
pursuing the Ph.D. degree in information and communications
engineering at the GIST. His research interests lie in the  broad
areas of compressed sensing, machine learning, wireless
communications, and statistical signal processing.
\end{IEEEbiography}
\begin{IEEEbiography}[{\includegraphics[width=1in,height=1.25in,clip,keepaspectratio]{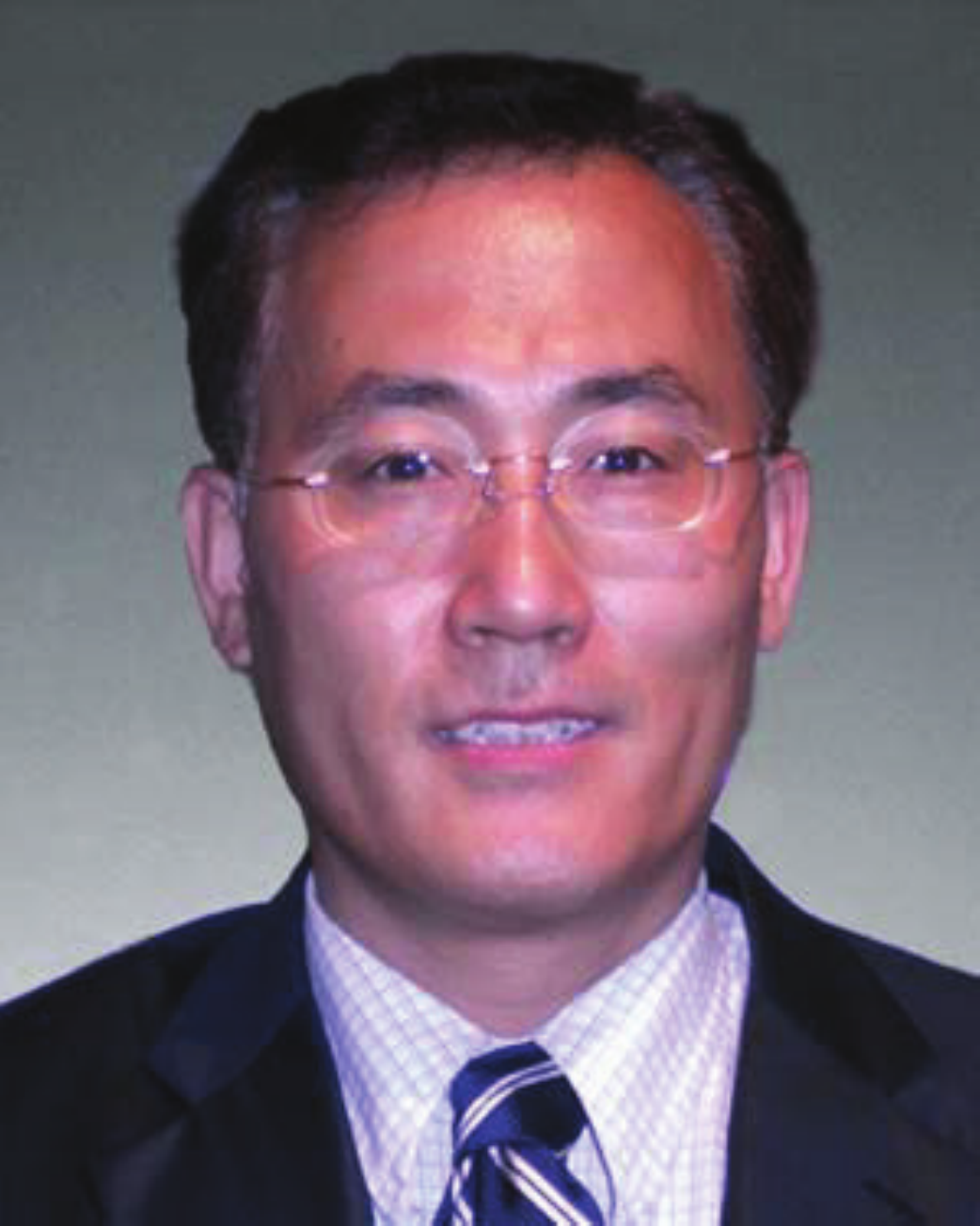}}]{Heung-No Lee}
(SM'13) received the B.S., M.S., and Ph.D. degrees from the
University of California, Los Angeles, CA, USA, in 1993, 1994, and
1999, respectively, all in electrical engineering. From 1999 to
2002, he was with  HRL Laboratories, LLC, Malibu, CA, USA, as a
Research Staff Member from 1999 to 2002. From 2002 to 2008, he was
with the University of Pittsburgh, Pittsburgh, PA, USA, as an
Assistant Professor. He joined  Gwangju Institute of Science and
Technology (GIST), Korea, where he is currently a Professor. His
general areas of research include information theory, signal
processing, communications/networking theory, and their application
to wireless communications and networking, compressive sensing,
future internet, and brain- computer interface.
\end{IEEEbiography}
\begin{IEEEbiography}[{\includegraphics[width=1in,height=1.25in,clip,keepaspectratio]{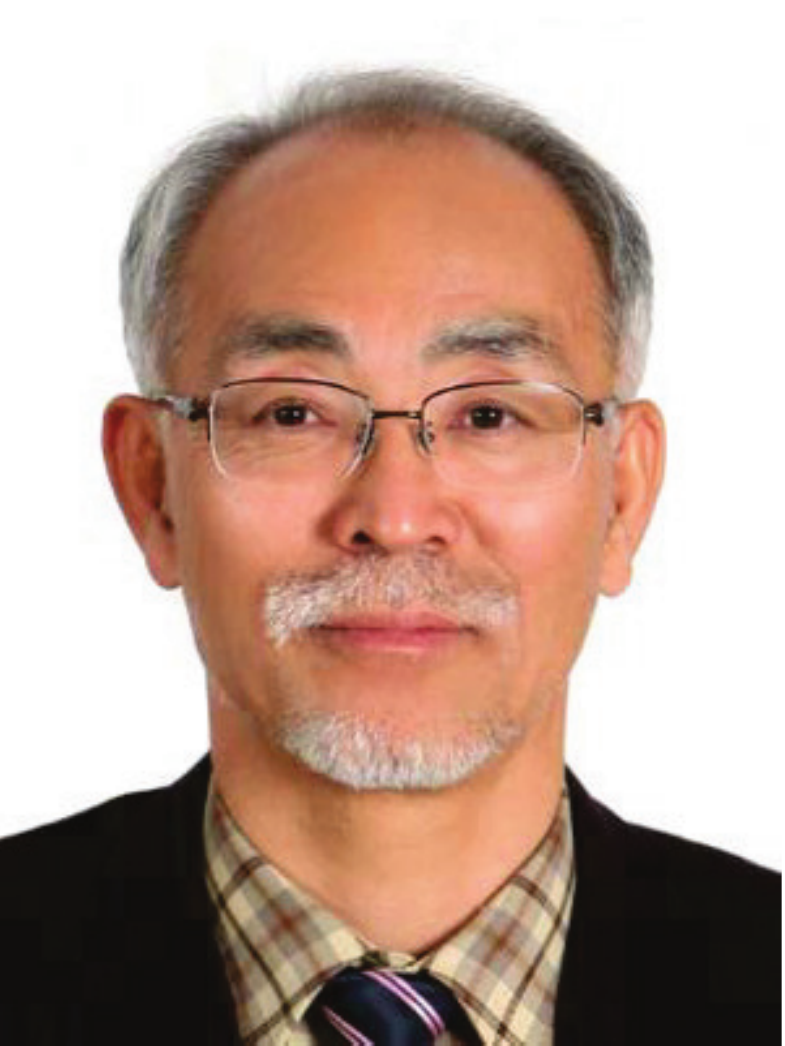}}]{Kiseon Kim}
(SM'98) received the B.Eng. and M.Eng. degrees, in electronics
engineering, from Seoul National University, Korea, in 1978 and
1980, and the Ph.D. degree in electrical engineering systems from
University of Southern California, Los Angeles, in 1987. From 1988
to 1991, he was with Schlumberger, Houston, Texas. From 1991 to
1994, he was with the Superconducting Super Collider Lab, Texas. He
joined Gwangju Institute of Science and Technology (GIST), Korea, in
1994, where he is currently a Professor. His current interests
include wideband digital communications system design, sensor
network design, analysis and implementation both, at the physical
layer and at the resource management layer.
\end{IEEEbiography}
\vfill


%



\begin{thebibliography}{10}



\bibitem{MRI} M. Lustig, D. L. Donoho, and J. M. Pauly,
``Sparse MRI: The application of compressed sensing for rapid MR
imaging," \emph{Magnetic Resonance in Medicine}, vol. 58, issue 6,
pp. 1182-1195, Dec. 2007.

\bibitem{SPC} M. Duarte, M. Davenport, D. Takhar, J. Laska, T. Sun, K. Kelly, and R. Baraniuk,
``Single-pixel imaging via compressive sampling,"  \emph{IEEE Signal
Processing Magazine}, vol. 25, no. 2, pp. 83-91, Mar. 2008,


\bibitem{ADC} M. Mishali, Y. C. Eldar, O. Dounaevsky, and E. Shoshan,
``Xampling: Analog to digital at sub-Nyquist rates," \emph{IET
Circuits, Devices and Systems}, vol. 5, issue 1, pp. 8-20, Jan.
2011.


\bibitem{Donoho2} D. L. Donoho, M. Elad, and V. Temlyakov, ``Stable
recovery of sparse overcomplete representations in the presence of
noise," \emph{IEEE Trans. Inform. Theory}, vol. 52, no. 1, pp. 6-18,
Jan. 2006.
%


\bibitem{tropp} J. A. Tropp, ``Just relax: convex programming methods for
identifying sparse signals in noise," \emph{IEEE Trans. Inform.
Theory}, vol. 52, no. 3, pp. 1030-1051, 2006.


\bibitem{candes2}  E. Candes and T. Tao, ``The Dantzig selector: statistical estimation when
p is much larger than n," \emph{Ann. Statist.}, vol. 35, no. 6, pp.
2313-2351, 2007.

\bibitem{Lasso} R. Tibshirani, ``Regression shrinkage and selection via the
lasso," \emph{J. Roy. Statisti. Soc., Ser. B}, vol. 58, no. 1, pp.
267-288, 1996.



%
\bibitem{OMP}  J. A. Tropp and A. C. Gilbert, ``Signal recovery from random
measurements via orthogonal matching pursuit," \emph{IEEE Trans.
Inform. Theory}, vol. 53, no. 12, pp. 4655-4666, Dec. 2007.

\bibitem{Cosamp} D. Needell, J. Tropp, "COSAMP: Iterative signal
recovery from incomplete and inaccurate samples," \emph{Appl. and
Comput. Harmon. Anal.}, vol. 26, no. 3, pp. 301-321, 2008.

\bibitem{Bruckstein} A. M. Bruckstein, D. L. Donoho, and M. Elad, ``From sparse solutions
of systems of equations to sparse modeling of signals and images,"
\emph{SIAM Rev.}, vol. 51, no. 1, pp. 34-81, Feb. 2009.

\bibitem{SBL} M. E. Tipping, ``Sparse Bayesian learning and the relevance vector
machine," \emph{J. Mach. Learn. Res.}, vol. 1, pp. 211-244, 2001.




\bibitem{BCS} Shihao Ji, Ya Xue, and Lawrence Carin, ``Bayesian compressive sensing,"
\emph{IEEE Trans. Signal Process.}, vol. 56, no. 6, pp. 2346-2356,
June. 2008.



%


%
%
%




\bibitem{CS-BP2} D. Baron, S. Sarvotham, and R. Baraniuk, ``Bayesian compressive sensing via belief propagation,"
 \emph{IEEE Trans. Signal Process.}, vol. 58, no. 1, pp. 269-280, Jan.
2010.

\bibitem{SSP2012} J. Kang, H.-N. Lee, and K. Kim, ``Bayesian hypothesis test for sparse support recovery using belief propagation,"
\emph{Proc. of IEEE Statistical Signal Processing Workshop (SSP)},
pp. 45-48, Aug. 2012.

\bibitem{BP-SBL} X. Tan and J. Li, ``Computationally efficient sparse Bayesian learning via belief propagation,"
\emph{IEEE Trans. Signal Process.}, vol. 58, no. 4, pp. 2010-2021,
Apr. 2010.

\bibitem{SuPrEM} M. Akcakaya, J. Park, and V. Tarokh, ``A coding
theory approach to noisy compressive sensing using low density
frame,"  \emph{IEEE Trans. Signal Process.,} vol. 59, no. 12, pp.
5369-5379, Nov. 2011.




\bibitem{AMP1} D. L. Donoho, A. Maleki, and A. Montanari, ``Message passing
algorithms for compressed sensing,", \emph{Proc. Nat. Acad. Sci.},
vol. 106, pp. 18914-18919, Nov. 2009.


%
%






\bibitem{factor} F. R. Kschischang, B. J. Frey, and H.-A. Loeliger, ``Factor
graphs and the sum-product algorithm," \emph{IEEE Trans. Inform.
Theory}, vol. 47, no. 2, pp. 498-519, Feb. 2001.




\bibitem{non_para_BP} E. Sudderth, A. Ihler, W. Freeman, and A. S. Willsky,
``Nonparametric belief propagation,"  \emph{Communi. of the ACM} vol
53, no. 10, pp. 95-103, Oct. 2010.

\bibitem{coughlan} J. M. Coughlan and S. J. Ferreira, ``Finding
deformable shapes using loopy belief propagation," \emph{Proc. of
12th Euro. Conf. on Comp. Vision (ECCV)}, pp. 453-468, 2002.

\bibitem{Isard} M. Isard, J. MacCormick, and K. Achan, ``Continuously-adaptive
discretization for message-passing algorithms," \emph{Proc. of the
Adv. in Neural Inform. Process. Sys. (NIPS)},  2009.

\bibitem{Noorshams} N. Noorshams, and M. J. Wainwright, ``Quantized stochastic belief propagation:
efficient message-passing for continuous state spaces," \emph{Proc.
of IEEE Int. Symp. Inform. Theory (ISIT)}, pp. 1246-1250, July,
2012.


\bibitem{nBP_sensor} A. T. Ihler, J. W. Fisher, R. L. Moses, and A. S. Willsky,
``Nonparametric belief propagation for self-localization of sensor
networks," \emph{IEEE Journal on Sel. Areas in Communi.}, vol 23,
no.4, pp. 809-819, Apr. 2005.




%
%
\bibitem{Gallager} R. G. Gallager, \emph{Low-Density Parity Check Codes},
MIT Press: Cambridge, MA, 1963.

\bibitem{Mackey} D. J. MacKay, \emph{Information theory, inference, and learning algorithms},
 Cambridge University Press, 2003, Available from www.inference.phy.cam.ac.uk/mackay/itila/

\bibitem{Richardson} T. Richardson, and R. Urbanke, ``The capacity of  low-density parity check
codes under message-passing decoding," \emph{IEEE Trans. Inform.
Theory}, vol. 47, no. 2, pp. 599-618, Feb. 2001.



\bibitem{Wainwright1} W. Wang, M. J. Wainwright, and K. Ramchandran, ``Information-theoretic
limits on sparse signal recovery: Dense versus sparse measurement
matrices," \emph{IEEE Trans. Inform. Theory}, vol. 56, no. 6, pp.
2967-2979, Jun. 2010.

\bibitem{Wainwright2}  M. J.  Wainwright, ``Information-theoretic limits on sparsity
recovery in the high-dimensional and noisy setting," \emph{IEEE
Trans. Inform. Theory}, vol. 55, no. 12, pp. 5728-5741, Dec. 2009.

\bibitem{Akcakaya} M. Akcakaya and V. Tarokh, ''Shannon-theoretic limit on noisy compressive sampling," \emph{IEEE Trans. Inform. Theory}, vol. 56, no. 1, pp.
492-504, Jan. 2010.

\bibitem{Fletcher} A. Fletcher, S. Rangan, and V. Goyal, ``Necessary and sufficient conditions for sparsity pattern recovery," \emph{IEEE Trans. Inform.
Theory}, vol. 55, no. 12, pp. 5758-5772, Dec. 2009.

\bibitem{DE_info} D. Middleton and R. Esposito, ``Simultaneous optimum detection and estimation of signal in noise," \emph{IEEE Trans. Inform.
Theory}, vol. 14, no. 3, pp. 434-444, May. 1968.




%
%

\bibitem{KayI} S. Kay, \emph{Fundamentals of Statistical Signal Processing Volume
I:Detection Theory}, Prentice Hall PTR, 1993.

\bibitem{KayII} S. Kay, \emph{Fundamentals of Statistical Signal Processing Volume
II:Estimation Thoery}, Prentice Hall PTR, 1993.

\bibitem{Bjorck} Ake Bjorck, \emph{Numerical Methods for Least
Squares Problems,},SIAM: PA, 1996.

\bibitem{sigmarule} F. Pukelsheim, ``The three sigma rule," \emph{The American Statistician}, vol. 48, no. 2, May. 1994.

\bibitem{spikeandslab} H. Ishwaran and J. S. Rao, ``Spike and slab variable selection :
Frequentist and Bayesian strategies," \emph{Ann. Statist.}, vol.33,
pp. 730-773, 2005.







\bibitem{Guo05} D. Guo and S. Verdu, ``Randomly spread CDMA:
Asymptotics via statistical physics," \emph{IEEE Trans. Inform.
Theory}, vol. 51, no. 6, pp. 1983-2010, Jun. 2005.

%
\bibitem{Guo08} D. Guo and C. C. Wang, ``Multiuser detection of sparsely spread
CDMA," \emph{IEEE J. Sel. Areas Comm.}, vol. 26, no. 3, pp. 421-431,
Mar. 2008.


\bibitem{Freg} Frey, B. J. and D. J. MacKay, ``A revolution:
Belief propagation in graphs with cycles,"  \emph{Proc. of the 11th
Annual Conference on Neural Inform. Proces. Sys., (NIPS)}, pp.
479-485, Dec. 1997.


\bibitem{loopyBP} K.P. Murphy,  Y.  Weiss, and M.I. Jordan, ``Loopy belief propagation for approximate inference: An empirical study,"
\emph{Proc. of the 5th conf. on Uncertainty in artificial
intelligence}, Morgan Kaufmann Publishers Inc., pp. 467-475, 1999.

\bibitem{Moustaki} G. Moustakides, G. Jajamovich, A. Tajer, and X.
Wang, ``Joint detection and estimation: optimum tests and
applications," \emph{IEEE Trans. Inform. Theory},  vol. 58, no. 7,
pp. 4215-4229, June 2012.

\bibitem{Bishop} C. M. Bishop, \emph{Pattern Recognition and Machine Learning},
Springer: NY, 2006.


\end{thebibliography}
\end{document}